\documentclass[twocolumn]{aastex631}

\usepackage{graphicx}
\usepackage{eurosym}
\usepackage{natbib}
\usepackage{booktabs}
\usepackage{amsmath}
\usepackage{graphics,graphicx}
\usepackage{hyperref}
\hypersetup{urlcolor=blue}
\usepackage{cleveref}
\usepackage{eurosym}
\usepackage{booktabs}
\usepackage{wrapfig}
\usepackage{tabularx}

\newcommand{\rsun}{R$_\odot$}
\newcommand\be{\begin{equation}}
\newcommand\ee{\end{equation}}

\def\ellq{l}
\def\dd{{\rm d}}

\def\PP{{\cal P}}

\shorttitle{Asphericity of the solar CZ base}
\shortauthors{Basu et al.}

\begin{document}

\title{Asphericity of the base of the solar convection zone}

\correspondingauthor{Sarbani Basu}
\email{sarbani.basu@yale.edu}

\author[0000-0002-6163-3472]{Sarbani Basu}
\affiliation{Department of Astronomy,
Yale University, PO Box 208101,
New Haven, CT 06520-8101,USA}

\author[0000-0003-1531-1541]{Sylvain G. Korzennik}
\affiliation{Center for Astrophysics $|$ Harvard \& Smithsonian, Cambridge, MA 02138, USA}

\begin{abstract}
We have used solar oscillation frequencies and frequency splittings  obtained over solar cycles 23 and 24 to investigate whether the base of the solar convection zone shows any departure from spherical symmetry. We used the even-order splitting coefficients, $a_2$--$a_8$, and estimated the contributions from each one separately. The average asphericity over the two solar cycles was determined using frequencies and splittings obtained with a 9216-day time-series. We find that evidence of asphericity is, {\em at best}, marginal: the $a_2$ component is consistent with no asphericity, the $a_4$ and $a_6$ components yield results at a level a little greater than $1\,\sigma$, while the $a_8$ component shows a signature below $1\,\sigma$. The combined results indicate that the time average of the departure from the spherically symmetric position of the base of the convection zone is $\lesssim 0.0001R_\odot$. We have also used helioseismic data  obtained from  time-series of lengths 360 days, 576 days, 1152 days, and 2304 days in order to examine the consistency of the results and evaluate whether there is any time variation. We find that the evidence for time variation is statistically marginal in all cases, except for the $a_6$ component, for which tests consistently yield $p$ values of less than $0.05$.
\end{abstract}


\section{Introduction} \label{sec:intro}

The base of the solar convection zone (henceforth CZ) marks the position where the thermal gradient changes from being radiative in the interior to adiabatic in the outer layers. The position where this change happens is known very well \citep[e.g.,][]{jcd1991, sba1997, basu1998}.
It is determined assuming that the base is spherically symmetric. In fact, the position of the CZ base is a test of the physics used in constructing solar models \citep[e.g.,][]{jcd1993}.

The base of the CZ is also where the solar tachocline is located. This is a narrow shear layer that marks the change from latitudinal differential rotation in the CZ to a solid body rotation in the interior. The tachocline however, is prolate in shape \citep{abc1998,paulchar1999}; at the equator this layer is centered below the CZ base,
while at a latitude of 60$^\circ$, it is above the CZ base. This raises the question, whether the base of the CZ is indeed spherically symmetric. \citet{zahn1992} noted that the tachocline could drive meridional flows. It has been argued \citep{dog_agk1995} that these flows must affect the thermal balance of the layers, and could influence where the layer becomes unstable to convection, thereby causing the base of the CZ to deviate from spherical symmetry. 

Rotation is expected to make the CZ base oblate, though the expected oblateness is small --- the surface value of the difference between equatorial and polar radius  is only a few parts in $10^6$ \citep{kuhn}, the rotation rate at the CZ base is not much different. 
Another expected cause of any asphericity of the CZ base, is the magnetic field that might be there. Indeed, there may be belts of magnetic fields intense enough to modify convection.  Once the magnetic fields become buoyant and rises up, convection  could be enhanced by the field,  thus the CZ base would exhibit time-dependent asphericity \citep[see][]{dog_agk1995}.

Previous investigations on the asphericity of the CZ base give results that do not agree with one another. 
Using frequencies obtained by the Big Bear Solar Observatory \citep{bbso},
\citet{dog_agk1995} used combinations of frequencies and splitting coefficients to determine the frequencies sensitive at three different latitudes, and inverted the resulting combinations to determine the relative isothermal difference between a spherically symmetric model and that implied by the frequency combinations. They interpreted their results as a
difference in the CZ base of no more than $0.02 R_\odot$ between the  pole and the equator. \citet{mario2001}  attempted to look for asphericity at the CZ base using data from the Michelson Doppler Imager \citep[MDI;][]{mdi}. They fitted the signature of the acoustic ``glitch'' caused by the change in the thermal gradient at the CZ base to determine the acoustic depth of this feature. They found a latitude-dependent difference in the acoustic depth, of the order of 100 seconds, and a small, time dependence at low latitudes. Like \citet{dog_agk1995}, they used combined frequencies and splittings to obtain frequencies that correspond to various latitudes, but their combinations were modified by an initial inversion to get localized latitudinal response of the combination. 
In contrast, \citet{me2001} found a much smaller difference, and derived an upper limit of $0.0005 R_\odot$ on the asphericity of the CZ base; 
they used data from the Global Oscillation Network Group \citep[GONG:][]{gong} and the MDI and, like \citet{dog_agk1995}, combined frequencies and splittings to obtain frequencies that
represent various latitudes. \citet{me2001} did not invert the frequencies, instead they used the technique developed by \citet{sba1997} to determine the position of the CZ base.
\citet{hma2001, hma2003} did two-dimensional inversions to determine the latitudinal dependence of solar structure, they found distinct asphericity in the outer layers, but nothing significant at the CZ base. 

Much more helioseismic data are now available, and in this paper, we use helioseismic data collected over solar cycles 23 and 24 to assess  whether the CZ base shows any asphericity and furthermore, whether there is any time variation of the asphericity. We use mode frequencies and splitting obtained by different projects, and more importantly, we use frequencies and frequency splittings obtained using different lengths of helioseismic time-series. Unlike all earlier attempts, we do not combine the frequencies and splittings to determine frequencies for a given latitude, but determine the contribution of each term separately and derive the uncertainty due to each term separately.

The rest of the paper is organized as follows. In Section~\ref{sec:asph} we describe how one can determine the combination of frequencies and frequency splittings in order to be sensitive to different latitudes, and in particular we describe our approach of determining the asphericity of the CZ base. In Section~\ref{sec:czbase}\ we describe the method of determining the position of the CZ base. Section~\ref{sec:data} describes the different data sets that we use. Our results are presented in Section~\ref{sec:res}, and our conclusions in Section~\ref{sec:conc}.

\section{Determining Asphericity} 
\label{sec:asph}

We follow the formulation presented in \citet{asph2001} and \citet{me2001}  to determine the asphericity at the base of the CZ. We use mode frequencies and their splittings, expressed in the usual manner as:
\begin{equation}
\nu_{nlm}
= \nu_{nl} + \sum_{j=1}^{j_{\rm max}} a_j (n,l) \, \PP_j^{(l)}(m), 
\label{eq:eq_split}
\end{equation}
where $\nu_{nl}$, or the central frequency, is determined by the spherically symmetric part of solar structure, $a_j$ are splitting coefficients and
$\PP_j$ are re-scaled Clebsch-Gordon coefficients \citep{SchouEtal:2002}. In this decomposition, the odd-order $a_j$ are caused  by the solar rotation, while the even-order coefficients contain the signature of asphericity and magnetic field.

It should be noted that with global helioseismic data, one cannot distinguish between magnetic fields and asphericity \citep[see e.g.,][]{zweibel}. We chose to interpret the signature in even-order splitting coefficients as being caused by asphericity, this is what has been done in all previous estimates of asphericity from global helioseismic data.

We consider only axisymmetric perturbations in structure,  with the
symmetry axis coinciding with the rotation axis. Since we use global modes, our inferences, like for the rotation rate, cannot differentiate between the northern and southern hemispheres, but return estimates for a mean hemisphere. 
This allows us to use the variational principle and write the difference in frequency between the Sun and a solar model for
a mode of a given order, spherical harmonic degree and azimuthal order ($n$, $\ellq$, and $m$) as:
\begin{multline}
{\delta\nu_{nlm}\over\nu_{nlm}}=
\int_0^R\;\dd r\;\int_0^{2\pi}\;\dd \phi\;\int_0^\pi \sin\vartheta\;\dd \vartheta\\
\left({\cal K}_{c^2,\rho}^{n\ellq}(r){\delta c^2\over c^2}(r,\vartheta)+
{\cal K}_{\rho,c^2}^{n\ellq}(r){\delta\rho\over\rho}(r,\vartheta)
\right)Y_\ellq^m(Y_\ellq^m)^*,
\label{eq:eqvar}
\end{multline}
where $r$ is radius, $\vartheta$ is the co-latitude,
${\delta\nu_{nlm}/\nu_{nlm}}$ is the relative frequency difference,
and  ${\cal K}_{c^2,\rho}^{n\ellq}(r)$ and ${\cal K}_{\rho,c^2}^{n\ellq}(r)$
are the kernels
for spherically symmetric perturbations \citep[see e.g.,][]{hmasb1994,sasha1999}. 
The functions $Y_\ellq^m$ are spherical harmonics, denoting the angular dependence of
the eigenfunctions for a spherically symmetric star, and are normalized such that 
\begin{equation}
\int_0^{2\pi}\;\dd\phi\;\int_0^\pi \sin\vartheta\;\dd\vartheta\;Y_\ellq^m(Y_\ellq^m)^* = 1. 
\label{eq:ylm}
\end{equation}

The angular integrals in Eq.~\ref{eq:eqvar} can be evaluated to give
\begin{multline}
\int_0^{2\pi}\;\dd\phi\;\int_0^\pi \sin\vartheta\;\dd\vartheta\;Y_\ellq^m(Y_\ellq^m)^*
P_{2k}(\cos\vartheta)=\\
{1\over\ellq}Q_{\ellq k}{\cal P}_{2k}^{(\ellq)}(m)
\label{eq:ang}
\end{multline}
where $Q_{\ellq k}$ depends only on $\ellq,k$ and ${\cal P}_{2k}^{(\ellq)}(m)$
are the orthogonal polynomials defined in Eq.~\ref{eq:eq_split}.
The extra factor of $1/\ellq$ ensures that $Q_{\ellq k}$ approach a constant
value at large $\ellq$. 

With this choice of expansion, the inversion problem can be decomposed
into independent inversions for each even splitting coefficient, $a_{2k}$, and
${\delta c^2/ c^2}$ and ${\delta\rho/\rho}$
can be expressed in terms of the even order Legendre polynomials, $P_i$:
\begin{equation}
{\delta c^2\over c^2}(r,\vartheta)=\sum_k ({\delta c^2\over c^2})_k(r) P_{2k}(\cos\vartheta), 
\label{eq:c2}
\end{equation}
and,
\begin{equation}
{\delta \rho\over \rho}(r,\vartheta)=\sum_k {(\delta\rho\over\rho})_k(r) P_{2k}(\cos\vartheta). 
\label{eq:rho}
\end{equation}

Thus, $({\delta c^2/ c^2})_k(r)$ and $({\delta\rho/\rho})_k(r)$ can be computed by inverting each of the splitting coefficient $a_{2k}$.
Note that this is similar to the ``1.5D'' inversions used to infer the rotation rate, where a two-dimensional solution is obtained
from a series of one dimensional inversions:
\begin{multline}
{\ellq a_{2k}{(n,\ellq)}\over\nu_{n\ellq}Q_{\ellq k}}=\\
\int_0^R{\cal K}_{c^2,\rho}^{n\ellq}c_k(r)\;\dd r+
\int_0^R{\cal K}_{\rho,c^2}^{n\ellq}\rho_k(r)\;\dd r.
\label{eq:1d}
\end{multline}

Instead of determining the CZ base at different colatitudes from the
corresponding frequency difference, we estimate the difference in CZ over the spherically symmetric value implied by the different orders of the even-order coefficients separately; we do this for $a_2$--$a_8$; higher order coefficient have large uncertainties and hence, we do not use those. In other words, we take the combination
\be
\delta\nu=\delta\nu_{n\ell}+{\ell a_{2k}(n,\ellq)\over Q_{\ell k}}
\label{eq:ak}
\ee
for $k=1,$2,3, and 4 separately, determine the position of the CZ base in each case, and then subtract the CZ base implied by $\delta\nu_{n\ell}$ to obtain
$\Delta R_{\rm CZ}(a_{2k})$ for each $a_{2k}$ under consideration. Adding $\delta\nu_{n\ell}$ is necessary for our method 
(see Section~\ref{sec:czbase}), which relies on determining the frequency differences between models with known CZ depth and the Sun to work properly. 
Once we have estimates of $\Delta R_{\rm CZ}$ for each case, they are combined to obtain the CZ base at any given co-latitude, i.e,
\be
R_{CZ}(\vartheta)=R_{\rm CZ}(\nu)+\sum_{k}\Delta R_{\rm CZ}(a_{2k})P_{2k}(\cos\vartheta),
\label{eq:comb}
\ee
where $R_{\rm CZ}(\nu)$ is the spherically symmetric part of the CZ base obtained using the central frequencies, and $\Delta R_{\rm CZ}(a_2k)$ is the contribution of a given even-order a-coefficient. We determine $\Delta R_{\rm CZ}(a_2k)$ for $k=1$,2,3 and 4 separately using the data combinations in Eq.~\ref{eq:ak}, and subtract the spherically symmetric part of the CZ base. We do not use higher-order $a$ coefficients since the uncertainties in those are very large, in fact, as we will see later in \S~\ref{sec:res}, the uncertainties even in $a_8$ are large enough to mask out any signs of asphericity. In the rest of the paper, when we talk of ``asphericity'' we refer to, depending on the context, either $\Delta R_{\rm CZ}(a_{2k})$ or $R_{CZ}(\vartheta)$. 

\begin{figure}[]
    \centering
    \includegraphics[width=3.3 true in]{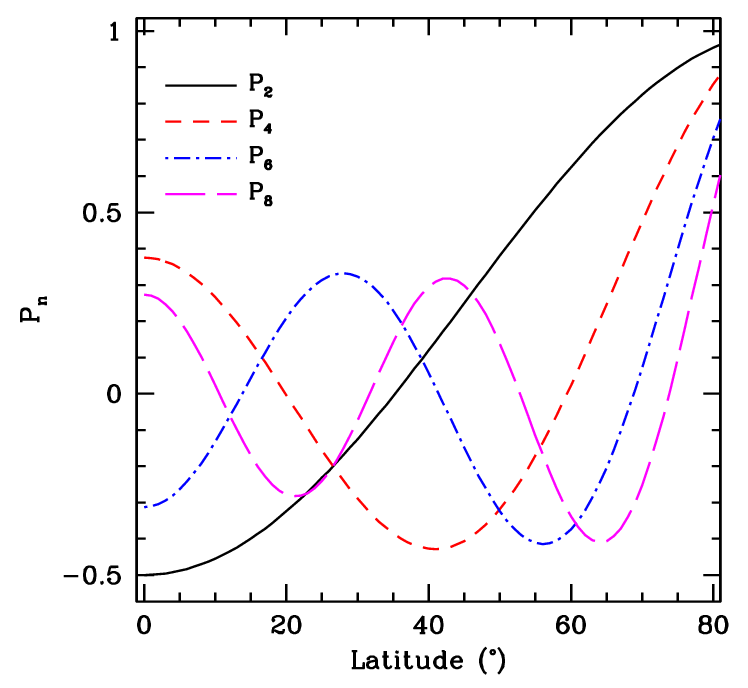}
    \caption{ $P_{2k}(\cos\vartheta)$ plotted as a function of latitude $\theta$, to show how different even-order $a$-coefficients contribute to the asphericity of the CZ base. Thus, for example, if $\Delta R_{\rm CZ}(a_2)$ is positive, the CZ base at the equator will be deeper than that at higher latitudes, making  it prolate. }
    \label{fig:legpol}
\end{figure}

In Fig.~\ref{fig:legpol}, we illustrate $P_{2k}$ as a function of latitude. The figure shows that a positive $\Delta R_{\rm CZ}(a_{2})$ implies that the CZ base
at the equator is deeper than that at the higher latitudes, namely that the CZ base is prolate. Of course, higher-order contributions make the shape more complex.

\section{Estimating the position of the CZ base}\label{sec:czbase}

\begin{figure}
    \centering
    \includegraphics[width=3.0 true in]{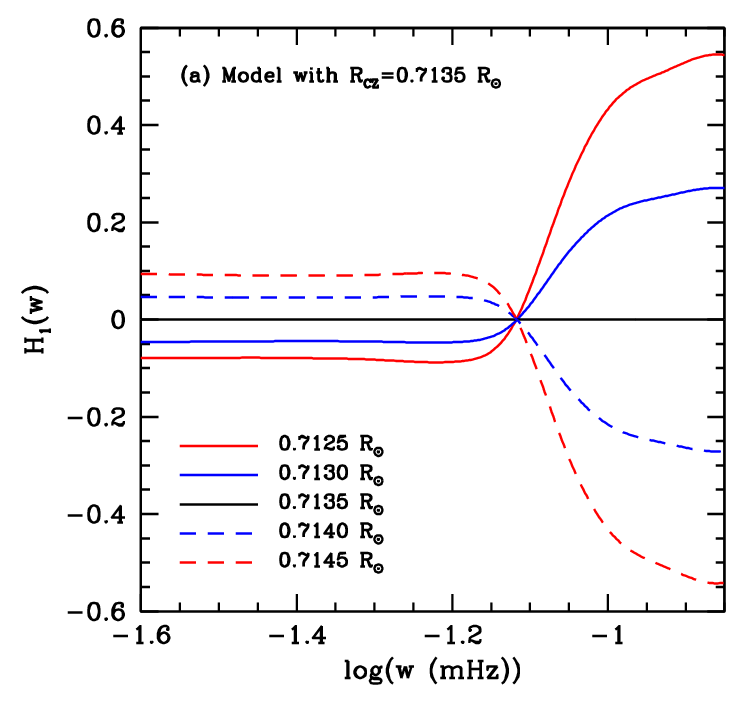}\\
    \includegraphics[width=3.0 true in]{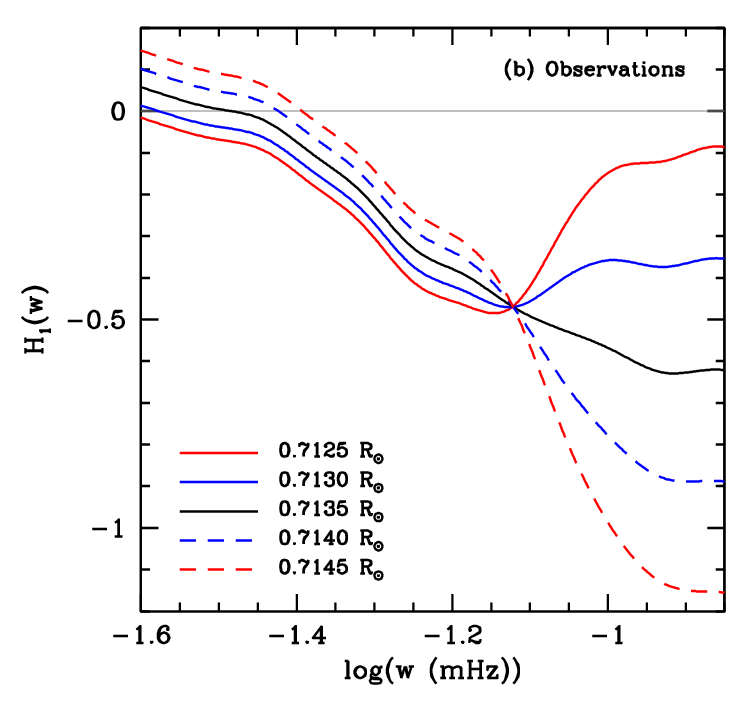}
    \caption{The function $H_1(w)$ between a (a) solar envelope model and (b) solar frequencies  and solar envelope models with different CZ positions. The test solar model in panel (a) has the same physics as the calibration models. The observational data used in panel (b) is an HMI set obtained with a 2304-day time-series with a start date of 2001.08.11. Note the need for the term $H_s(w)$  in Eq.~\ref{eq:calib} in order to fit the curves in panel (b) with those in panel (a).}
    \label{fig:h1h2}
\end{figure}

We use the method developed and used by \citet{sba1997,basu1998} to determine the position of the CZ base, 
that can be summarized as follows. 
If there are two otherwise similar
solar models with different depths of the convection zone, then
just below the base of the convection zone
the model with a deeper convection zone will have a larger
sound speed than the other.
This observable difference of sound speeds can be calibrated to find the
convection zone depth of a test model or the Sun, as was first done by \citet{jcd1991}. 

Asymptotically, the frequency differences between a solar model
and the Sun, or between two solar models can be written as \citep{cz1989}
\be
S(w){\delta \omega\over \omega}=H_1(w)+H_2(\omega),
\label{eq:diff}
\ee
where
$w=\omega/(\ellq+0.5)$, $\omega = 2\pi\nu$, and  $S(w)$ is a known function for a given
reference model.
The functions $H_1(w)$ and $H_2(\omega)$ can be determined by a least-squares fit to the known frequency differences. The function $H_2$ contains the contribution from the ``surface term,'' i.e., the contribution from our inability to model the near-surface layers of the Sun properly. $H_1(w)$ can be inverted to obtain the sound-speed difference, $\delta c/c$, between the reference model and the Sun.  However, \citet{sba1997} showed that that is not required, since $H_1(w)$ itself can be used to determine the convection-zone depth.

If $\phi(w)$ is the $H_1(w)$ between two solar models which differ only in the depth of the convection zone, then $H_1(w)$ for any other pair of models can be written as
\be
H_1(w)=\beta\phi(w)+H_s(w),
\label{eq:calib}
\ee
where $H_s(w)$ is a smooth component of $H_1(w)$ which
results from sound-speed differences that arise from differences
in the equation of state, abundances, surface physics etc.,
and  the first term is the contribution to $H_1(w)$ due to the
sound-speed difference caused by the difference in $r_b$, the
position of the base of the CZ. Thus, if $\beta$ is
estimated using a least-squares fit, the unknown $r_b$ of the Sun can
be obtained.
We determine $\beta$ for a series of calibration models with
different $r_b$, and interpolate to find the CZ position for which $\beta=0$. This ``null'' method allows us to determine the position of the base of the CZ to a precision that is much better than what can be obtained with a sound-speed inversion, and is not hampered by the limited spatial resolution of 
inversions at the CZ base.

The uncertainty in our results  arising from data uncertainties is
evaluated by Monte Carlo simulations. For each observed data set, we create 4000 sets of artificial data by adding random realizations of the observed uncertainty to the central frequency and splitting coefficients. We determine $R_{\rm eff}$ and $\Delta R_{\rm eff}$ for each of the simulated sets. The final parameters that we adopt is the mean of the distribution of results and the standard deviation is the uncertainty. We have found that the distribution is Gaussian and hence using the mean and standard deviation is  adequate.

As is clear from the above discussion, we need to use models with a specified
CZ depth. This can be done if we construct  models of the solar envelope,
for which the position of the CZ base and the helium abundance of the solar
envelope can be specified.

\citet{sba1997} had shown that the largest source of systematic error in the estimated CZ-base position is the hydrogen abundance profile. This is because the sound speed near the base of the CZ is affected not only by the change in the temperature gradient but also by the change in the mean molecular
weight due to gravitational settling of helium. This error does not affect this work since we are
looking at differences with respect to the spherically symmetric component of the CZ base, and thus this ought to cancel out. Nevertheless, to minimize the error, we use models that have 
$X$ and $Z$ profiles for the Sun determined by \citet{hmasmc1998}. However, since the position of the CZ base is well-known, we use calibration models with a smaller range in CZ positions --- 0.7125\rsun, 0.7130\rsun, 0.7135\rsun, 0.7140\rsun, and 0.7145\rsun. 
We show in Fig.~\ref{fig:h1h2} how $H_1(w)$ changes when solar models with different CZ positions are used.

\section{Data used}
\label{sec:data}

\begin{figure}
    \centering
    \includegraphics[width=0.50\textwidth]{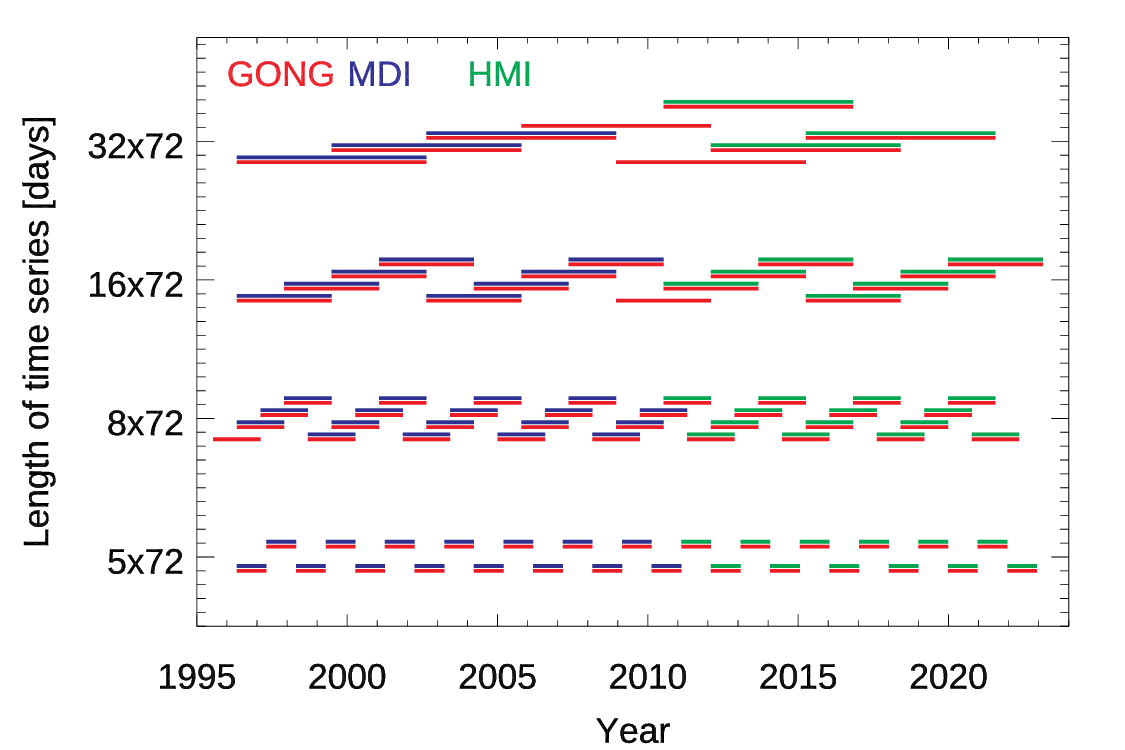}
    \caption{A visual representation of the time periods spanned by each SGK data set. Sets have been offset vertically for clarity. The sets, from top to bottom, are for $32\times72$, $16\times72$, $8\times72$ and $5\times72$ days respectively. Red denotes GONG, blue MDI and green HMI observations.}
    \label{fig:times}
\end{figure}

For this work, we use solar oscillation frequencies obtained by the ground-based  Global Oscillation Network Group \citep[GONG;][]{gong} and the space-based Michelson Doppler Imager (MDI) on board the Solar and Heliospheric Observatory spacecraft \citep{mdi} and the Helioseismic and Magnetic Imager \citep[HMI;][]{hmi}  on board the Solar Dynamics Observatory. We use mode frequencies obtained from different lengths of the time-series that cover solar cycles 23 and 24.  In particular, we use
data analyzed by an alternative pipeline \citep{syl1, syl2, syl3, syl4, syl5, sgk2018, sgk2023}, independent of the project's pipeline, and henceforth referred to as the ``SGK'' pipeline. The mode parameters are derived from time-series that are multiples of 72 days, we use mode frequencies and splittings obtained for time-series lengths of $32 \times, 16 \times, 8 \times$ and $5 \times 72$ days, or $\simeq 6, 3, 1.5$ or $1$ year, respectively. 
{The choice of analyzing time series that are multiples of 72 days was dictated by the standard time-series length adopted by the MDI team, and was later adopted by the HMI team. A time-series of 72 days was considered to be long enough to produce a frequency spectrum with high signal-to-noise ratio but
short enough to maintain ``good'' temporal variation, since the mode frequencies
were known to vary with time on all time-scales. The MDI and HMI teams later started analyzing 360-day (i.e., $5\times 72$-day) long time series too.} 

Fig.~\ref{fig:times} shows the temporal ranges the different data sets span. {For time-series longer than 72 days, the convention adopted by SGK is to fit
time-series separated by half the length of the time-series to increase the
temporal resolution. The exception is for the 360-day long time-series, these are fitted every 360 day, to match what is done by the MDI project.} Additionally, to determine the average asphericity over the two solar cycles, we used a GONG data set resulting from fitting a $128\times72$-day time-series.

\begin{figure}[h]
    \centering
    \includegraphics[width=3.35 true in]{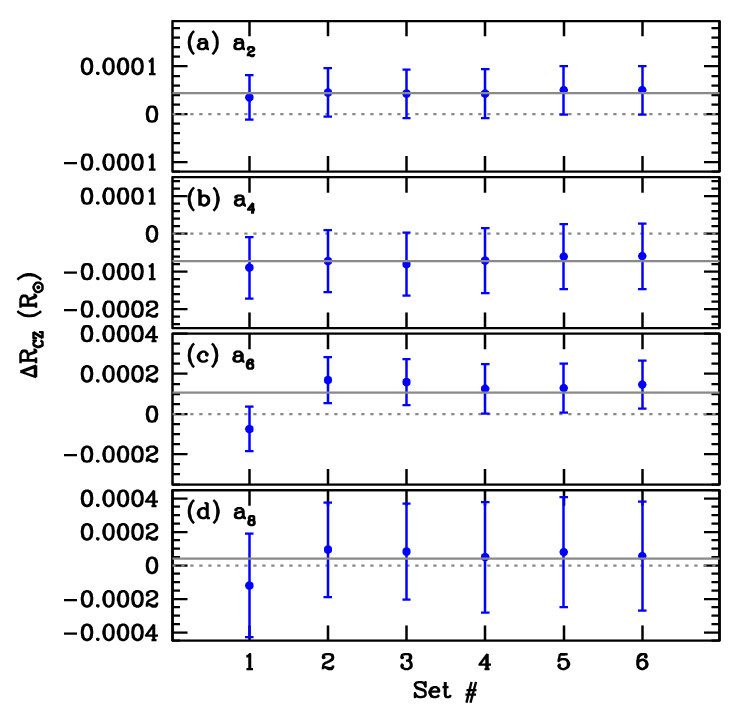}
    \caption{The figure shows the effect of the leakage matrix on $\Delta R_{CZ}$ implied by even-order splitting coefficients $a_2$--$a_8$ (Panels (a), (b), (c) and (d) respectively) obtained for mode frequencies and splittings obtained with $32\times72$-day time-series data from GONG and HMI with a start date of 2010:07:11 
    using different leakage matrices.  The solid gray horizontal line in each panel is the weighted mean of all the results. The dotted gray line marks where results would lie if the CZ base were spherically symmetric. Set 1 is the HMI pipeline set, while set 2 is the SGK pipeline using GONG data and the JSBo=0 leakage matrix, set 3 is the SGK pipeline using GONG and the SKBo=0 leakage matrix, set 4 is the SGK pipeline using the HMI data and the JSBo=0 leakage matrix, set 5 is the SGK pipeline using HMI data and the  SKBo=0 leakage matrix, and set 6 is the SGK pipeline using the HMI data and the SKBo=a2 leakage matrix.  }
    \label{fig:leakage}
\end{figure}

\begin{figure*}[htb]
    \centering
    \includegraphics[width=3.0 true in]{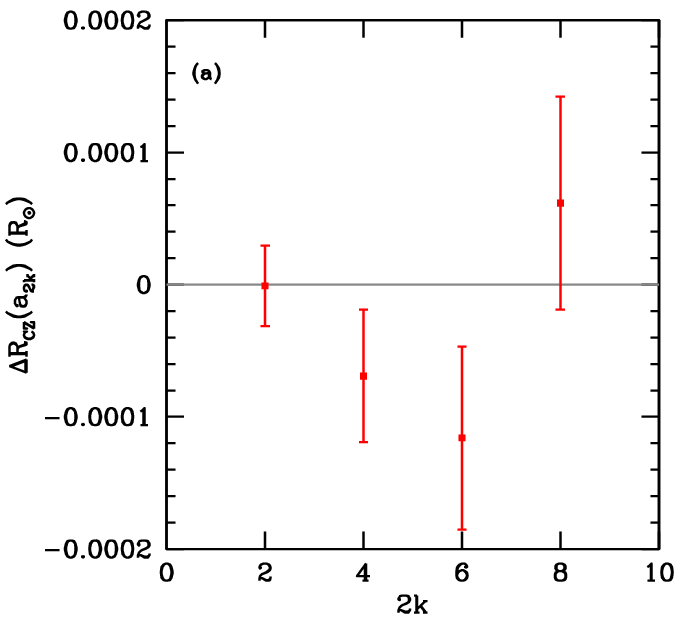}
    \includegraphics[width=3.0 true in]{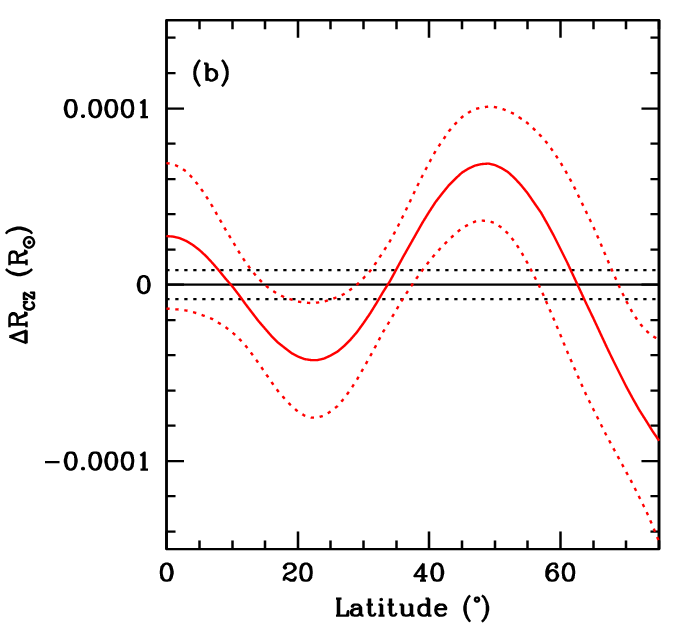}
    \caption{Left: Deviations from sphericity of the base of the CZ implied by different splitting coefficients $a_{2k}$ for $2k=1$--4. Right: The latitudinal variation of the asphericity implied by the results of the individual coefficients is shown as the red continuous line with the dotted line showing $1\sigma$ uncertainties. The black dotted lines show the $1\sigma$ uncertainty limit for the spherically symmetric component of the CZ base. }
    \label{fig:128e}
\end{figure*}

In some case, we also use mode parameters from the MDI and HMI projects' pipeline. {Although} the primary  frequency sets derived by these projects are obtained by fitting 72-day long time-series, which is not adequate for our work. However, there are some sets of frequencies and splittings obtained fitting $5\times 72$-day long time-series, and also one HMI set obtained from fitting a $32\times 72$-day long time-series. 

It should be noted that the SGK pipeline predominantly obtains mode parameters by fitting asymmetric profiles to peaks in the power spectra, though some sets fitted using symmetric profiles are available. The MDI and HMI projects' pipeline predominantly fits symmetric profiles, though some sets fitted using asymmetric profiles are available. 

We restrict ourselves to modes with frequencies between 1.6 mHz and 3.5 mHz that have lower turning points between 0.45\rsun\ and 0.95\rsun. These modes give good coverage of the CZ base, while keeping mode uncertainties low.

The SGK pipeline produces modes and splittings using a number of different leakage matrices. In one case, it uses the leakage matrix computed by the MDI/HMI project (JSBo=0), in the other cases it uses a leakage matrix computed by Korzennik (SKBo=0 \& SKBo=a2). Such leakage matrix coefficients are computed as follows: a series of images of spherical harmonics on the unit sphere are computed, using the observational image projection, for either a heliographic latitude at disk center, $B_o$ of 0, or $5.06^\circ$, the quadratic mean of $B_o$ over a year. Each image is then spatially decomposed using the same decomposition procedure that is used for the respective observations. However, as can be seen in Fig.~\ref{fig:leakage}, the leakage matrix does not change the results to any level of statistical significance. The results in Fig.~\ref{fig:leakage} are for $32\times72$-day data sets with a start date of 2010:07:11. Unfortunately, MDI data does not exist for that time period. However, tests with MDI data for other spans of time give very similar results. Consequently, for all subsequent results reported in this paper, we only use modes obtained with the JSBo=0 leakage matrix.

\begin{figure*}[hb]
    \centering
    \includegraphics[width=3.35 true in]{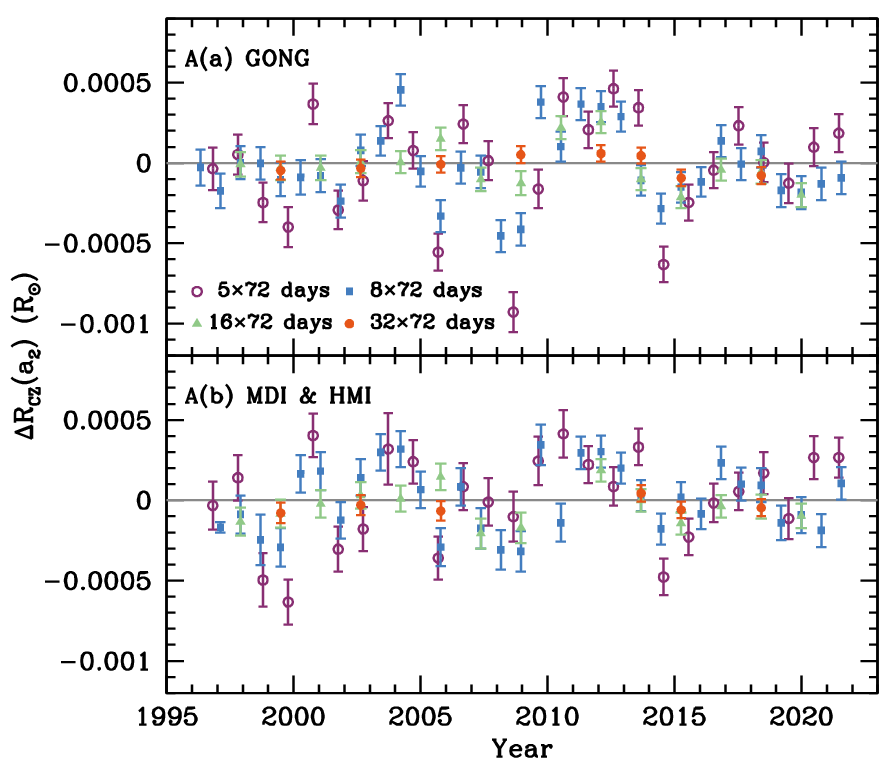}\includegraphics[width=3.35 true in]{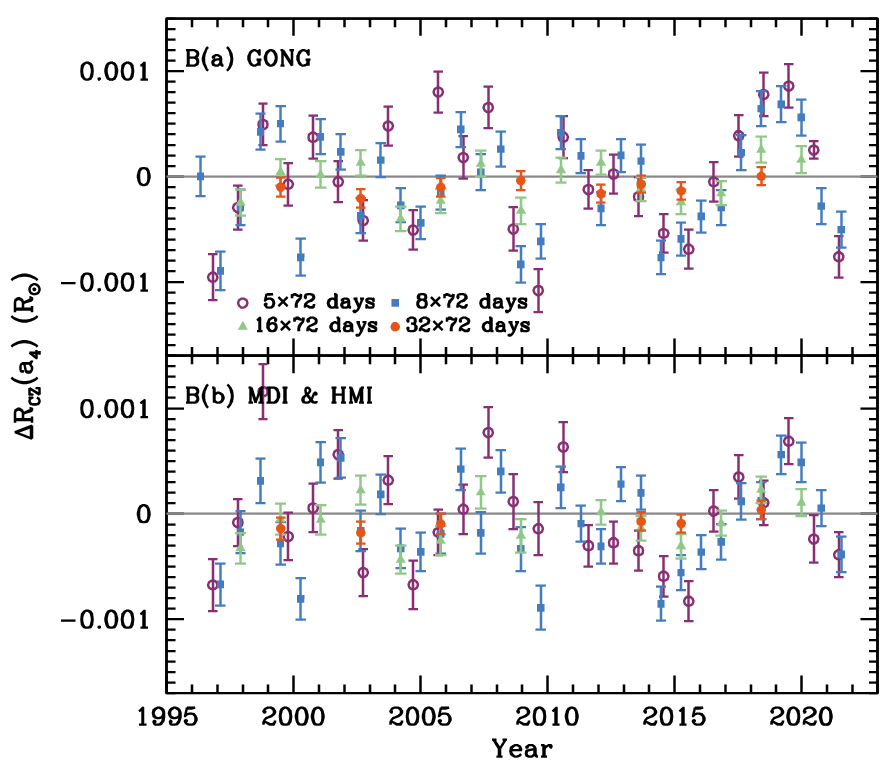}\\
    \includegraphics[width=3.35 true in]{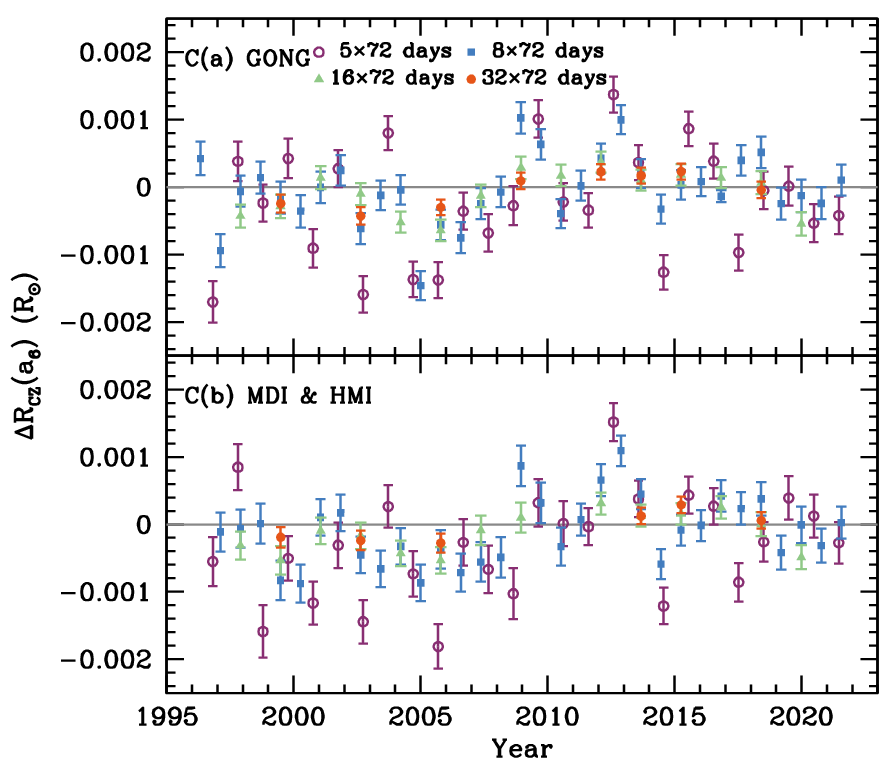}\includegraphics[width=3.35 true in]{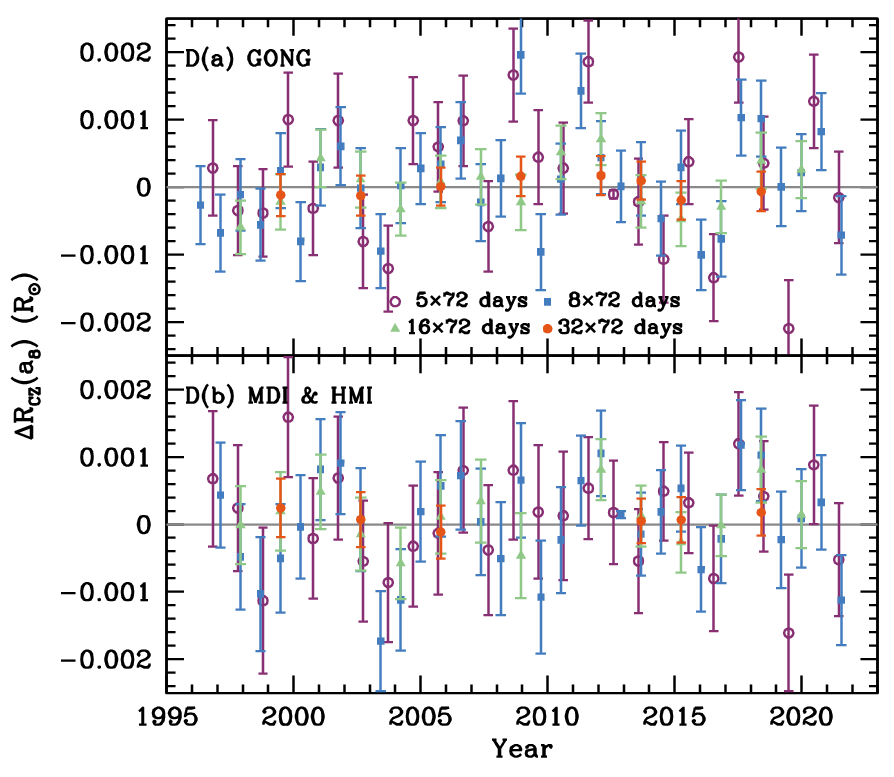}
    \caption{The deviation of the CZ base from spherical, $\Delta R_{\rm CZ}$, obtained with different lengths of time-series plotted as a function of time. Panels (A)--(D) are results for $a_2$-$a_8$ respectively. Sub-panel (a) in all panels shows results with GONG data, while sub-panel (b) are for MDI and HMI data. Only results obtained with the SGK pipeline fitting asymmetric profiles are shown. 
    }
    \label{fig:mainres}
\end{figure*}

\begin{figure*}[]
    \centering
    \includegraphics[width=6.00 true in]{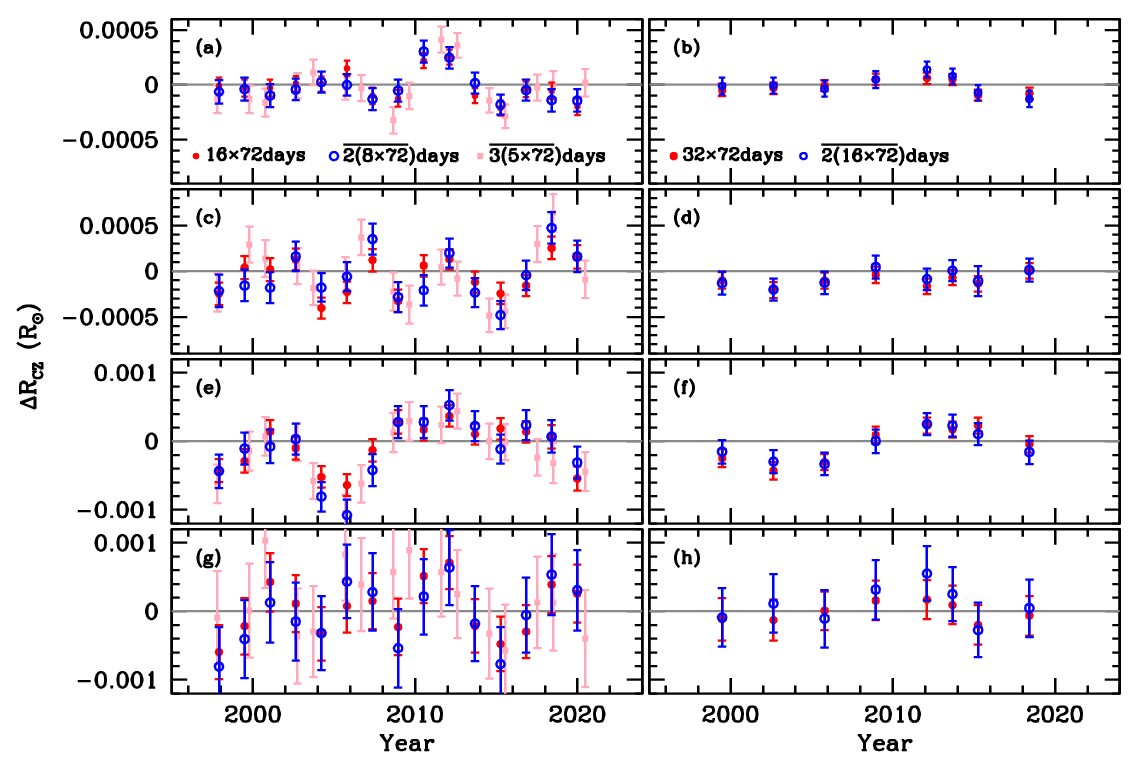}
    \caption{A comparison of our estimates of $\Delta R_{\rm CZ}$ obtained with long time-series data and those
    with averaged mode parameters of shorter data. Only results obtained with GONG data are shown, since the data cover both solar cycle 23 and 24. The column on the left compares results from $16\times72$-day time-series with those obtained with average mode parameters of 2 $(8\times 72)$-day sets and 3 $(5\times 72)$-day sets. The column on the right compares $32\times72$-day results with those obtained from the average mode parameters of 2 $(16\times 72)$-day time-series. The rows from top to bottom show results for $a_2$-$a_8$ respectively.}
    \label{fig:timeav}
\end{figure*}

\section{Results} \label{sec:res}

Our results derived from the $128\times72$-day set are shown in Fig.~\ref{fig:128e}. The values are tabulated in Table~\ref{tab:128e} in the Appendix. We find that the departure from sphericity, averaged over solar cycles 23 and 24 is extremely small, with only marginal statistical significance. When the contributions of the different components are combined to give the latitudinal difference, it is clear that we can at most give an upper limit of $0.0001R_\odot$.

Earlier investigations, mentioned in Section~\ref{sec:intro} obtained much larger deviations. This could have been because of the quality of the then available data, or possibly because there is a time dependence to the asphericity, which is averaged out by the long, $128\times72$-day, time-series. 
To test this, we used data sets obtained from shorter time-series. The resulting estimates of $\Delta R_{\rm CZ}$ are shown in  Fig.~\ref{fig:mainres}, and are tabulated in the Appendix.

A few features can be seen immediately. The scatter in the results obtained with the shorter time-series data is larger than those with the longer time-series data. This is, of course, expected given that the uncertainty in
the mode frequencies and splittings is lower for the longer sets. The results obtained with GONG data agree with those of MDI and HMI data. The agreement is, however, better between GONG and HMI than between GONG and MDI.
This type of disagreement has been seen in other helioseismic investigations, even when using the respective projects' pipeline data \citep[e.g.,][]{nssl}.

We can also see that the scatter in the results do not appear to be strictly random.
{The right-hand panels of Fig.~\ref{fig:timeav} give the impression of a possible anti-correlation between solar activity and CZ asphericity. Tests for such anti-correlation, where we used the 10.7cm radio flux averaged over the same time interval as an index of solar activity, do not reveal any statistically significant correlation. The largest anti-correlation is obtained for the $a_4$ results obtained with $32\times72$-day time series, and even that is significant only at the 2-$\sigma$ level.  }
{A strict solar cycle dependence is not really expected --- the time variation of the tachocline is, after all, much more complicated with a clear sign of hysteresis \citep{sbhma2019}, hence there may still be a }
time variation {just like in the tachocline}.
This seeming time variation is clearer for the results derived from the longer time-series, however, the variation becomes smaller as the length of the time-series is increased. 
This is to be expected if the time variation is real, since the longer sets would average out the time variations, 
but this can be merely a result of the observed uncertainties rather than a real temporal variation --- as the uncertainties get reduced, the amplitude of the inferred variations will decrease.

We first verify whether the results obtained by the shorter time-series are consistent with those obtained with longer time-series. For this, we compare the results obtained with $16 \times 72$-day data with those obtained with the average mode parameters of shorter sets --- we averaged mode frequencies and splittings of 2 $8\times72$-day sets and 3 $5\times 72$-day sets for each $16 \times 72$-day set, we also averaged mode frequencies and splittings of 2 $16\times72$-day sets for each $32\times72$-day set. 
Note that 3 $5\times 72$-day sets do not quite cover a $16\times 72$-day set, but they are close enough. 
Also note that obtaining mode parameters from a long time-series is not the same as averaging mode parameters from shorter time-series because in the former case many more modes can rise above the noise level, while in the latter the set of successfully fitted modes is not always the same. The results are shown in Fig.~\ref{fig:timeav}. We also show results obtained by averaging frequencies and splittings of 2 $16\times72$-day sets and compare those with the results of the $32\times 72$-day sets. As can be seen, averaging the mode parameters of the shorter sets yields results that are commensurate with those obtained with the longer sets. 

Figs.~\ref{fig:mainres} and \ref{fig:timeav} also shows that while $a_2$, $a_4$ and, $a_6$ results show deviations from asphericity $> 1\,\sigma$ at some epochs, while the uncertainties in the $a_8$ results are too large to make any firm deduction about the contribution of $a_8$ to the asphericity of the CZ base.  

\subsection{Tests for time variations}

The different panels in Figs.~\ref{fig:mainres} suggest the possibility of a time variation; such putative variation is clearer in Fig.~\ref{fig:timeav}. 

Given that the deviations from zero are small, we have performed additional tests to ascertain whether there really is a non-random variation. 

We performed a Wald–Wolfowitz runs test \citep{ww}. While we tested the results from the GONG data sets separately, we combine the MDI and HMI sets. 
However, since our results have uncertainties, the results of one runs test is not adequate. Thus, for each data set (i.e., each length of time-series), we generated multiple sets by adding random realizations of the uncertainties to the results at each epoch.  We also create a truly random set of results at each epoch; these are just random realizations of the uncertainties in the results. For each sequence of the realization of results and as well as the random set, we perform the runs test to derive a distribution of runs. 

We compare the distribution of the runs test from our results and the random sequence. 
Note that since the uncertainties in the results at each epoch are very similar, the distribution of runs in our random case can be described by a Gaussian. 
We conduct the usual two-sample Kolmogorov-Smirnov test, as well as a Student's t~test and an F~test, to examine the null hypothesis that the distribution of runs in our results is drawn from the distribution of runs for the sample whose time variation is random. 
Because we are essentially dealing with small number statistics (in terms of the number of epochs covered, in other words the number of runs we can get), 
we relied on multiple measures of differences between the distributions of runs in the results and the distribution of runs in the random sample.

The statistical tests yield marginal evidence for a non-random time variation in most cases. The tests generally yield $p$ values greater than 0.05 for $a_2$ and $a_4$, however, for $a_6$, we consistently get $p$ values less than $0.05$, and in some cases even more significant evidence of non-randomness.  The combination of MDI and HMI results show the same pattern, though the significance is generally lower.

\subsection{Examining possible sources of error}
\label{subsec:syst}

As shown above, the time-averaged signature of asphericity is very small, evidence of time-variation is also marginal. 
To ensure that our results are robust,  we take a close look at possible sources of error that could cause a spurious time dependence in our results.

\subsubsection{The second-order effects of rotation}

The even-order $a$ coefficients have a small contribution from the second-order effects of rotation \citep{gt1990}. It is known that solar rotation changes with time
\citep[see e.g.,][etc.]{antiabasu2013, howeetal2013, rudi2018, rachel2018, sasha2019, nssl}, even near the base of the solar convection zone \citep{sbhma2019}, the region of interest for this work. We did not correct the data for the second-order effect of rotation. 
In order to examine the effect of the second-order effects, we remove those from the GONG $16\times72$-day data-sets; we do so using the formalism of \citet{antia2000} and determine the asphericity again.

\begin{figure}[h]
    \centering
    \includegraphics[width=3.35 true in]{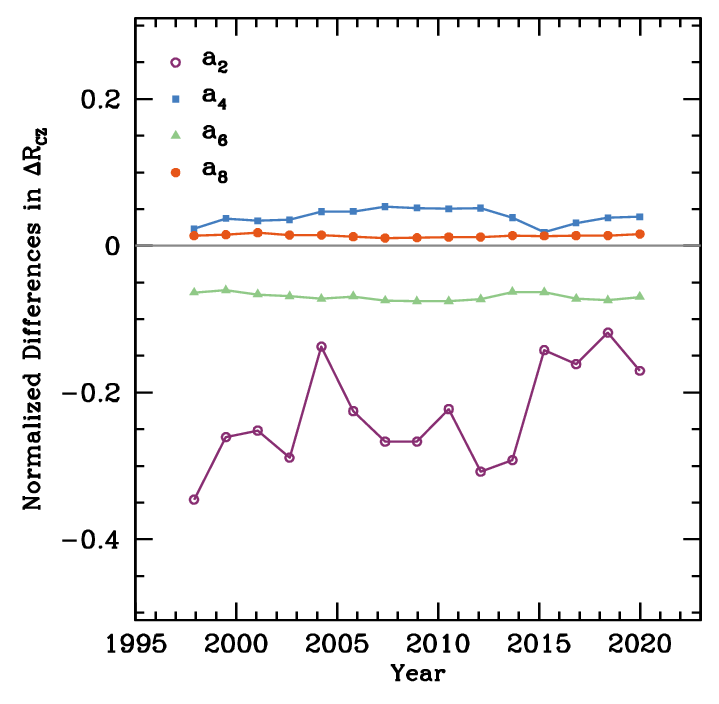}
    \caption{The difference in the original estimates of asphericity and those estimated using data sets corrected for the second order effects of rotation, normalized by the uncertainties in the original result. These results were obtained with GONG $16\times72$-day data sets.}
    \label{fig:2nd}
\end{figure}

The differences between the uncorrected and corrected results, normalized by the uncertainty in the corrected result, are shown in Fig.~\ref{fig:2nd}. As can be seen, while the differences are systematic, they are a small fraction of the uncertainty. The difference is largest for $\Delta R_{\rm CZ}(a_2)$, but even there, the differences are less  than $0.5\,\sigma$. Furthermore, although rotation is time-dependent, that does not affect the asphericity results.

\subsubsection{The surface term}

\begin{figure}[h]
    \centering
    \includegraphics[width=3.35 true in]{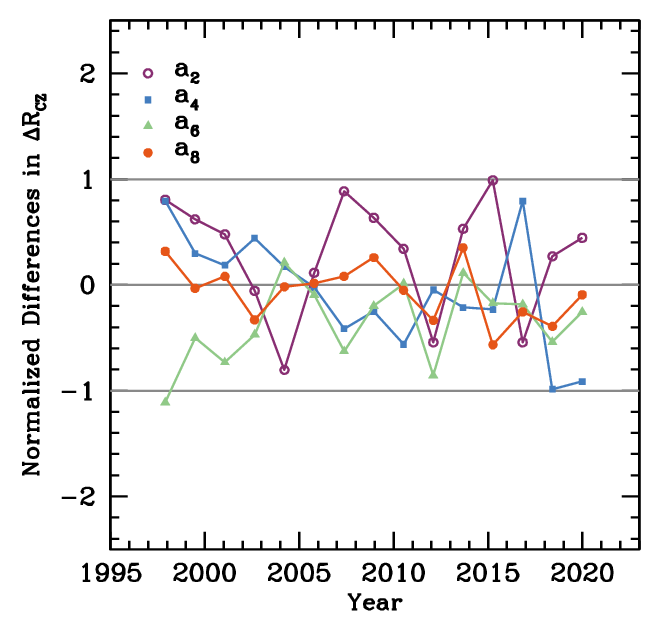}
    \caption{The change in $\Delta R_{\rm cz}$ estimates, when the splittings were restricted to modes
    with frequencies $\le 3$ mHz, normalized by their uncertainty, obtained when using $16\times72$-day GONG data sets. The horizontal lines mark $0$ and $\pm 1\,\sigma$ differences. Note that most results are well within $1\,\sigma$ and that there is no solar-cycle dependence.}
    \label{fig:surface}
\end{figure}

The surface term is the frequency-dependent frequency offset between the Sun and a solar model. This offset increases with frequency. It is well known that the solar surface term changes with solar cycle \citep[][etc.]{elsworth,libbrecht,rachel}, being largest at solar maximum and smallest at the minimum. The surface term has a steep frequency difference, and hence, to examine whether we remove it completely through Eq.~\ref{eq:diff}, we re-determine $\Delta R_{\rm CZ}$ after reducing the
upper frequency limit from 3.5 to 3 mHz. 
The results are shown in Fig.~\ref{fig:surface}.
As can be seen, differences between results with different upper frequency cut-off are generally well within $1\,\sigma$. These differences do not show a solar cycle dependence, something we would expect to see had we not removed the surface term.

\subsection{Asymmetry of the mode profiles}

\begin{figure}[]
    \centering
    \includegraphics[width=3.35 true in]{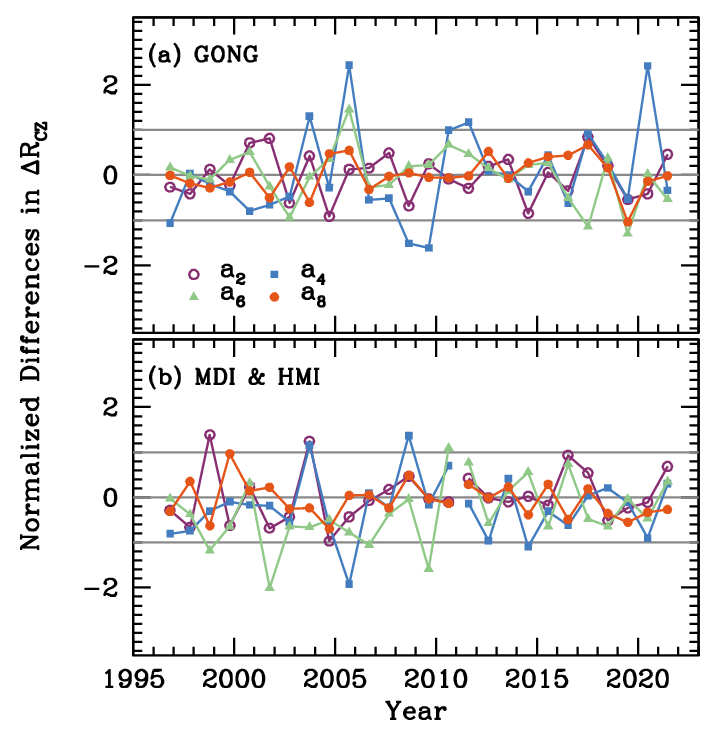}
    \caption{The difference in $\Delta R_{\rm cz}$ estimates, normalized by uncertainty, obtained with mode parameters asymmetric and symmetric fits to the data from the SGK pipeline for the $5\times72$-day time-series. The horizontal lines mark $0$ and
    $\pm 1\,\sigma$ differences.}
    \label{fig:symasym}
\end{figure}

As mentioned earlier, the data sets that we use determined frequencies and splittings by fitting an asymmetric profile to the peaks in the oscillation power spectra. In order to discount asymmetry as  the source of time variation, we compared the $5\times72$-day results obtained when fitting using either symmetric or asymmetric mode profiles. The differences, normalized by the uncertainty, are shown in Fig.~\ref{fig:symasym}. As can be seen, the differences are generally within $1\sigma$ and the time variations do not look anything like the ones seen in Figs.~\ref{fig:mainres} and \ref{fig:timeav}.

\begin{figure}[]
    \centering
    \includegraphics[width=3.35 true in]{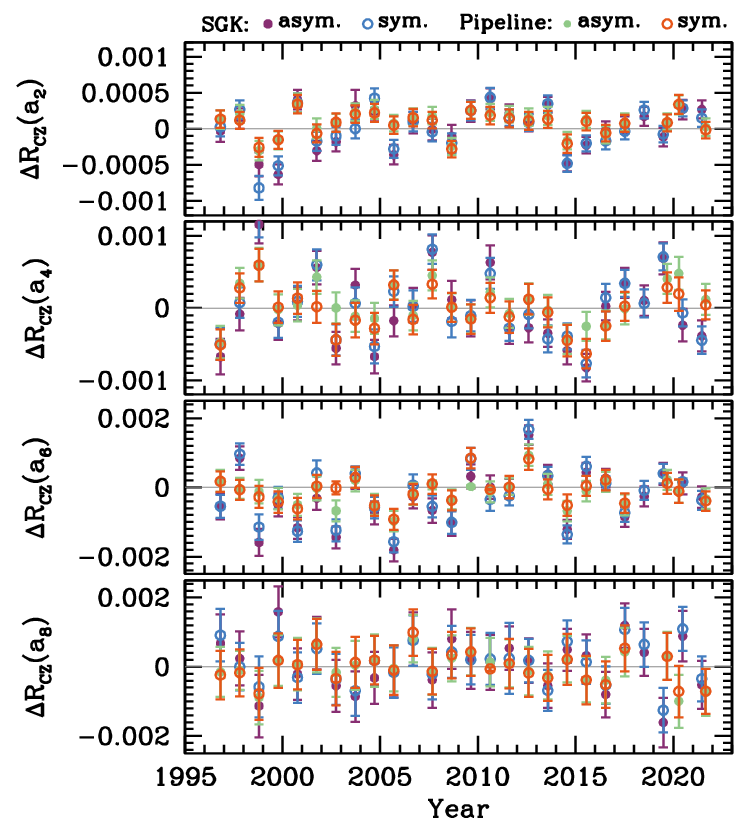}
    \caption{Asphericity parameter $\Delta R_{\rm CZ}$ obtained with $5\times72$-day MDI and HMI project pipeline data as well as with SGK pipeline data fitted using both symmetric and asymmetric peak profiles. The rows from top to bottom are for $a_2$--$a_8$ respectively. While Fig.~\ref{fig:symasym} shows the normalized difference between two sets, this figure shows the actual values for four sets of data.}
    \label{fig:comp}
\end{figure}

The MDI and HMI project pipelines also produce data sets from $5\times72$-day time-series mainly by fitting symmetric profiles, but some sets have also been fit using asymmetric profile. This allows us to compare results obtained with the project pipeline data with the SGK pipeline data, as shown in Fig.~\ref{fig:comp}.

\section{Discussion and Conclusions}
\label{sec:conc}

We have used global helioseismic data, frequencies and $a_2$--$a_8$ even-order splitting coefficients, 
available from various sources and derived using different methodologies as well as time-series of various lengths, all collected over solar cycles 23 and 24, to determine whether the base of the solar convection zone is aspherical. 

Using a data set obtained from a $128\times72$-day time-series, we find that the time-averaged asphericity of the CZ base is small, with an upper limit of $0.0001R_\odot$ on the deviation from sphericity. 

Our new estimate is smaller than those derived by earlier work. There are a number of reasons this could be the case. 
For one, all previous estimates used data sets from much shorter time-series, which means larger uncertainties that results in a larger uncertainty on the amplitude. 
Alternatively, if there is indeed a small variation, using a shorter time-series could make it more visible. And yet another possible reason for the discrepancy is the method by which asphericity is determined.

Our results are much smaller than the estimates of \citet{dog_agk1995}. It is possible that the \citet{dog_agk1995} inversion results (for the isothermal sound speed $u$ with the helium abundance $Y$ as the second parameter) have errors because calculating ($u,Y$) kernels requires one to assume that the equation of state of solar material is known and is the same as that of the reference model; \citet{sbjcd1997} have shown that this leads to systematic errors in the inversion results both above and below the CZ base and also results in unrealistically small uncertainties. 

It is more difficult to compare our results with those of \citet{mario2001} since they do not translate their acoustic depth differences into differences in radius. An acoustic depth difference of 100 seconds at the CZ base translates to a radius difference of approximately 0.025 R$_\odot$, again much larger than what we find. One of the possible reasons for this is that \citet{mario2001} only fit the acoustic glitch at the CZ base and neglect the contribution from  the helium ionization zone, and additionally do not take the surface-term into account. The glitch signature from the helium ionization zone is known to vary with solar cycle \citep{mandel, watson}, and the surface term is not only time dependent, the time-dependence depends on latitude \citep{asph2001, bab2007}. Not taking the latitudinal dependence of the surface term into account could lead to a large asphericity, while the time-varying signature from the helium ionization zone could add to the time-dependence of their results.
Our results are smaller than, but in line with, those of \citet{me2001}.
\citet{me2001}  used shorter time-series data and averaged the results over  slightly more than four years, but that is still a much shorter time interval than 25-years, which is what we use for our main results.

We have tested our results to examine whether there is a non-random time dependence in the results. Only the $a_6$ component gives a $p$-value of less than $0.05$. 
Our efforts to determine whether there is a real time-dependence are hampered by the small number of data sets we have.  
Results obtained with data from short time-series (like the usual 72-day sets from the project pipelines)  have unfortunately very large uncertainties, 
but increasing the length of the time-series decreases the number of sets covering the two solar cycles, and additionally average out any underlying time dependence.

Determining the cause of the asphericity, and any possible changes thereof, is beyond the scope of the paper. However, assuming that the asphericity is an indirect result of magnetic fields and the resulting modification of the thermal structure, a shift in the CZ base of $0.0001R_\odot$ implies a magnetic field of $\sim 300$ kG. This is consistent with the limits on the magnetic fields at the base of the CZ of a few hundred kilo-Gauss \citep[e.g.,][]{dsilva1993, me1997, antia2000, hma2003}.

\section*{Acknowledgements}
We thank H.~M.~Antia for his comments on an earlier draft of the paper that led us to conduct statistical tests to ascertain the significance of the temporal variations and for his help in calculating the second-order effects of rotation.
This work is supported by NASA grant 80NSSC23K0563 to SB and NASA grants 80NSSC22K0516 and  NNH18ZDA001N-DRIVE to SGK.
This work utilizes data from the National Solar Observatory Integrated Synoptic Program, which is operated by the Association of Universities for Research in Astronomy, under a cooperative agreement with the National Science Foundation and with additional financial support from the National Oceanic and Atmospheric Administration, the National Aeronautics and Space Administration, and the United States Air Force. The GONG network of instruments is hosted by the Big Bear Solar Observatory, High Altitude Observatory, Learmonth Solar Observatory, Udaipur Solar Observatory, Instituto de Astrof\'{\i}sica de Canarias, and Cerro Tololo Interamerican Observatory.
This work also uses provided by the SOHO/MDI consortium. SOHO is a project of international cooperation between ESA and NASA. We also use data from the Helioseismic and Magnetic Imager on board the Solar Dynamics Observatory. HMI data are courtesy of NASA/SDO and the AIA, EVE, and HMI science teams.


\appendix

We tabulate our results for the SGK asymmetric sets in Tables~\ref{tab:128e}, \ref{tab:32e}, \ref{tab:16e}, \ref{tab:8e} and \ref{tab:5e}. Note that the start date of each set is in YYYYMMDD format. 

\begin{table*}[h]
  \centering
  \caption{Results for the $128\times72$-day set}
  \scriptsize
\begin{tabular}{lcccc}
\toprule
Start Date &\multicolumn{4}{c}{$\Delta R_{\rm CZ}$/R$_\odot$} \\
of set & $a_2$ & $a_4$ & $a_6$ & $a_8$ \\
\toprule
19960501 & $-$9.4590E$-$07 $\pm$ 3.032E$-$05  &
$-$6.913E$-$05 $\pm$ 5.007E$-$05 &
$-$1.159E$-$04 $\pm$ 6.921E$-$05 &  
\phantom{$-$}6.173E$-$05 $\pm$ 8.067E$-$05\\
\toprule
\end{tabular}
\label{tab:128e}
\end{table*}

\begin{table*}[h]
  \centering
  \caption{Results for the $32\times72$-day sets}
  \scriptsize
\begin{tabular}{lcccc}
\toprule
Start Date &\multicolumn{4}{c}{$\Delta R_{\rm CZ}$/R$_\odot$} \\
of set & $a_2$ & $a_4$ & $a_6$ & $a_8$ \\
\hline
 & \multicolumn{4}{c}{GONG}\\
\hline 
19960501&  $-$4.738E$-$05           $\pm$    5.718E$-$05 &   $-$1.002E$-$04            $\pm$   9.185E$-$05 &   $-$2.444E$-$04            $\pm$   1.333E$-$04 &   $-$1.163E$-$04 $\pm $    3.116E$-$04\\
  19990627&  $-$3.263E$-$05           $\pm$    5.430E$-$05 &   $-$2.082E$-$04            $\pm$   8.657E$-$05 &   $-$4.283E$-$04            $\pm$   1.279E$-$04 &   $-$1.265E$-$04 $\pm $    2.975E$-$04\\
  20020822&  $-$8.575E$-$06           $\pm$    5.061E$-$05 &   $-$1.016E$-$04            $\pm$   8.703E$-$05 &   $-$3.006E$-$04            $\pm$   1.172E$-$04 &   \phantom{$-$}8.281E$-$06 $\pm $    2.824E$-$04\\
  20051017&  \phantom{$-$}5.070E$-$05 $\pm$    5.268E$-$05 &   $-$3.764E$-$05            $\pm$   9.057E$-$05 &   \phantom{$-$}9.180E$-$05  $\pm$   1.219E$-$04 &   \phantom{$-$}1.601E$-$04 $\pm $    2.911E$-$04\\
  20081212&  \phantom{$-$}5.876E$-$05 $\pm$    5.171E$-$05 &   $-$1.637E$-$04            $\pm$   8.471E$-$05 &   \phantom{$-$}2.272E$-$04  $\pm$   1.165E$-$04 &   \phantom{$-$}1.718E$-$04 $\pm $    2.881E$-$04\\
  20100711&  \phantom{$-$}4.541E$-$05 $\pm$    5.081E$-$05 &   $-$7.202E$-$05            $\pm$   8.235E$-$05 &   \phantom{$-$}1.693E$-$04  $\pm$   1.152E$-$04 &   \phantom{$-$}9.342E$-$05 $\pm $    2.830E$-$04\\
  20120207&  $-$9.347E$-$05           $\pm$    5.085E$-$05 &   $-$1.363E$-$04            $\pm$   8.488E$-$05 &   \phantom{$-$}2.298E$-$04  $\pm$   1.161E$-$04 &   $-$1.971E$-$04 $\pm $    2.869E$-$04\\
  20150404&  $-$7.865E$-$05           $\pm$    5.255E$-$05 &   \phantom{$-$}3.237E$-$06  $\pm$   8.627E$-$05 &   $-$4.353E$-$05            $\pm$   1.204E$-$04 &   $-$6.526E$-$05 $\pm $    2.908E$-$04\\
\hline
 & \multicolumn{4}{c}{MDI}\\
\hline
 19960501&  $-$7.867E$-$05           $\pm$    6.447E$-$05 &   $-$1.385E$-$04            $\pm$   1.051E$-$04 &   $-$1.907E$-$04            $\pm$   1.518E$-$04 &   \phantom{$-$}2.438E$-$04 $\pm $    4.312E$-$04\\
  19990627&  $-$2.981E$-$05           $\pm$    6.308E$-$05 &   $-$1.783E$-$04            $\pm$   1.025E$-$04 &   $-$2.384E$-$04            $\pm$   1.443E$-$04 &   \phantom{$-$}7.186E$-$05 $\pm $    4.077E$-$04\\
  20020822&  $-$6.684E$-$05           $\pm$    6.030E$-$05 &   $-$9.508E$-$05            $\pm$   9.902E$-$05 &   $-$2.753E$-$04            $\pm$   1.419E$-$04 &   $-$1.137E$-$04 $\pm $    3.954E$-$04\\
\hline
 & \multicolumn{4}{c}{HMI}\\
\hline
 20100711&  \phantom{$-$}4.329E$-$05 $\pm$    5.089E$-$05 &   $-$7.144E$-$05            $\pm$   8.622E$-$05 &   \phantom{$-$}1.255E$-$04  $\pm$   1.221E$-$04 &   \phantom{$-$}4.889E$-$05 $\pm $    3.299E$-$04\\
  20120207&  $-$5.962E$-$05           $\pm$    5.164E$-$05 &   $-$9.135E$-$05            $\pm$   8.593E$-$05 &   \phantom{$-$}2.892E$-$04  $\pm$   1.213E$-$04 &   \phantom{$-$}6.755E$-$05 $\pm $    3.368E$-$04\\
  20150404&  $-$4.604E$-$05           $\pm$    5.262E$-$05 &   \phantom{$-$}3.790E$-$05  $\pm$   8.924E$-$05 &   \phantom{$-$}5.797E$-$05  $\pm$   1.240E$-$04 &   \phantom{$-$}1.780E$-$04 $\pm $    3.445E$-$04\\
\toprule
\end{tabular}
\label{tab:32e}
\end{table*}

\begin{table*}[h]
  \centering
  \caption{Results for the $16\times72$-day sets}
  \scriptsize
\begin{tabular}{lcccc}
\toprule
Start Date &\multicolumn{4}{c}{$\Delta R_{\rm CZ}$/R$_\odot$} \\
of set & $a_2$ & $a_4$ & $a_6$ & $a_8$ \\
\hline
 & \multicolumn{4}{c}{GONG}\\
\hline 
19960501 & $-$9.578E$-$06  $\pm$   7.559E$-$05   & $-$2.472E$-$04  $\pm$   1.247E$-$04   & $-$4.292E$-$04  $\pm$   1.695E$-$04   & $-$5.980E$-$04  $\pm$   3.937E$-$04\\
  19971128 & $-$3.080E$-$05  $\pm$   7.665E$-$05   &  \phantom{$-$}4.067E$-$05  $\pm$   1.254E$-$04   & $-$2.863E$-$04  $\pm$   1.721E$-$04   & $-$2.182E$-$04  $\pm$   4.119E$-$04\\
  19990627 & $-$3.039E$-$05  $\pm$   7.607E$-$05   &  \phantom{$-$}1.888E$-$05  $\pm$   1.256E$-$04   &  \phantom{$-$}1.395E$-$04  $\pm$   1.717E$-$04   &  \phantom{$-$}4.238E$-$04  $\pm$   4.263E$-$04\\
  20010123 &  \phantom{$-$}8.732E$-$06  $\pm$   7.203E$-$05   &  \phantom{$-$}1.294E$-$04  $\pm$   1.196E$-$04   & $-$1.026E$-$04  $\pm$   1.630E$-$04   &  \phantom{$-$}1.114E$-$04  $\pm$   4.160E$-$04\\
  20020822 &  \phantom{$-$}4.753E$-$06  $\pm$   6.905E$-$05   & $-$4.011E$-$04  $\pm$   1.150E$-$04   & $-$5.198E$-$04  $\pm$   1.572E$-$04   & $-$3.296E$-$04  $\pm$   3.920E$-$04\\
  20040320 &  \phantom{$-$}1.501E$-$04  $\pm$   7.010E$-$05   & $-$2.298E$-$04  $\pm$   1.168E$-$04   & $-$6.399E$-$04  $\pm$   1.588E$-$04   &  \phantom{$-$}7.852E$-$05  $\pm$   3.892E$-$04\\
  20051017 & $-$1.015E$-$04  $\pm$   7.376E$-$05   &  \phantom{$-$}1.206E$-$04  $\pm$   1.230E$-$04   & $-$1.313E$-$04  $\pm$   1.660E$-$04   &  \phantom{$-$}1.512E$-$04  $\pm$   4.082E$-$04\\
  20070516 & $-$1.252E$-$04  $\pm$   7.478E$-$05   & $-$3.250E$-$04  $\pm$   1.248E$-$04   &  \phantom{$-$}2.841E$-$04  $\pm$   1.684E$-$04   & $-$2.288E$-$04  $\pm$   4.118E$-$04\\
  20081212 &  \phantom{$-$}2.202E$-$04  $\pm$   7.036E$-$05   &  \phantom{$-$}6.091E$-$05  $\pm$   1.172E$-$04   &  \phantom{$-$}1.707E$-$04  $\pm$   1.603E$-$04   &  \phantom{$-$}5.158E$-$04  $\pm$   3.947E$-$04\\
  20100711 &  \phantom{$-$}2.534E$-$04  $\pm$   6.866E$-$05   &  \phantom{$-$}1.315E$-$04  $\pm$   1.133E$-$04   &  \phantom{$-$}3.695E$-$04  $\pm$   1.567E$-$04   &  \phantom{$-$}7.099E$-$04  $\pm$   3.859E$-$04\\
  20120207 & $-$1.009E$-$04  $\pm$   6.826E$-$05   & $-$1.183E$-$04  $\pm$   1.133E$-$04   &  \phantom{$-$}1.037E$-$04  $\pm$   1.557E$-$04   & $-$2.133E$-$04  $\pm$   3.893E$-$04\\
  20130905 & $-$2.129E$-$04  $\pm$   6.937E$-$05   & $-$2.424E$-$04  $\pm$   1.162E$-$04   &  \phantom{$-$}1.825E$-$04  $\pm$   1.587E$-$04   & $-$4.772E$-$04  $\pm$   3.979E$-$04\\
  20150404 & $-$4.178E$-$05  $\pm$   6.884E$-$05   & $-$1.570E$-$04  $\pm$   1.153E$-$04   &  \phantom{$-$}1.379E$-$04  $\pm$   1.571E$-$04   & $-$2.948E$-$04  $\pm$   3.890E$-$04\\
  20161031 & $-$5.547E$-$05  $\pm$   7.416E$-$05   &  \phantom{$-$}2.548E$-$04  $\pm$   1.240E$-$04   &  \phantom{$-$}6.474E$-$05  $\pm$   1.689E$-$04   &  \phantom{$-$}3.919E$-$04  $\pm$   4.142E$-$04\\
  20180530 & $-$1.997E$-$04  $\pm$   7.548E$-$05   &  \phantom{$-$}1.588E$-$04  $\pm$   1.278E$-$04   & $-$5.430E$-$04  $\pm$   1.733E$-$04   &  \phantom{$-$}2.581E$-$04  $\pm$   4.238E$-$04\\
\hline
 & \multicolumn{4}{c}{MDI}\\
\hline
 19960501 & $-$1.333E$-$04  $\pm$   8.700E$-$05   & $-$3.274E$-$04  $\pm$   1.439E$-$04   & $-$3.134E$-$04  $\pm$   2.064E$-$04   & $-$1.311E$-$05  $\pm$   5.777E$-$04\\
  19971128 & $-$8.289E$-$05  $\pm$   8.848E$-$05   & $-$5.004E$-$05  $\pm$   1.460E$-$04   & $-$5.322E$-$04  $\pm$   2.104E$-$04   &  \phantom{$-$}1.935E$-$04  $\pm$   5.817E$-$04\\
  19990627 & $-$2.209E$-$05  $\pm$   8.436E$-$05   & $-$5.543E$-$05  $\pm$   1.396E$-$04   & $-$9.628E$-$05  $\pm$   1.983E$-$04   &  \phantom{$-$}4.837E$-$04  $\pm$   5.509E$-$04\\
  20010123 &  \phantom{$-$}3.061E$-$05  $\pm$   8.254E$-$05   &  \phantom{$-$}2.247E$-$04  $\pm$   1.365E$-$04   & $-$1.705E$-$04  $\pm$   1.933E$-$04   & $-$1.451E$-$04  $\pm$   5.396E$-$04\\
  20020822 &  \phantom{$-$}9.667E$-$06  $\pm$   8.087E$-$05   & $-$4.345E$-$04  $\pm$   1.344E$-$04   & $-$4.311E$-$04  $\pm$   1.901E$-$04   & $-$5.783E$-$04  $\pm$   5.295E$-$04\\
  20040320 &  \phantom{$-$}1.466E$-$04  $\pm$   8.305E$-$05   & $-$2.539E$-$04  $\pm$   1.373E$-$04   & $-$5.356E$-$04  $\pm$   1.957E$-$04   &  \phantom{$-$}1.133E$-$04  $\pm$   5.460E$-$04\\
  20051017 & $-$2.067E$-$04  $\pm$   9.270E$-$05   &  \phantom{$-$}2.023E$-$04  $\pm$   1.536E$-$04   & $-$8.969E$-$05  $\pm$   2.179E$-$04   &  \phantom{$-$}3.453E$-$04  $\pm$   6.127E$-$04\\
  20070516 & $-$1.721E$-$04  $\pm$   9.550E$-$05   & $-$2.101E$-$04  $\pm$   1.586E$-$04   &  \phantom{$-$}1.006E$-$04  $\pm$   2.252E$-$04   & $-$4.621E$-$04  $\pm$   6.324E$-$04\\
\hline
 & \multicolumn{4}{c}{HMI}\\
\hline
 20100711 &  \phantom{$-$}1.855E$-$04  $\pm$   6.960E$-$05   &  \phantom{$-$}1.310E$-$05  $\pm$   1.160E$-$04   &  \phantom{$-$}3.097E$-$04  $\pm$   1.631E$-$04   &  \phantom{$-$}8.094E$-$04  $\pm$   4.522E$-$04\\
  20120207 &  \phantom{$-$}8.665E$-$07  $\pm$   7.001E$-$05   & $-$1.380E$-$04  $\pm$   1.169E$-$04   &  \phantom{$-$}1.291E$-$04  $\pm$   1.629E$-$04   &  \phantom{$-$}1.249E$-$04  $\pm$   4.519E$-$04\\
  20130905 & $-$1.444E$-$04  $\pm$   6.895E$-$05   & $-$3.080E$-$04  $\pm$   1.159E$-$04   &  \phantom{$-$}1.568E$-$04  $\pm$   1.612E$-$04   & $-$2.675E$-$04  $\pm$   4.466E$-$04\\
  20150404 & $-$3.823E$-$05  $\pm$   6.993E$-$05   & $-$8.838E$-$05  $\pm$   1.177E$-$04   &  \phantom{$-$}2.527E$-$04  $\pm$   1.646E$-$04   & $-$1.293E$-$05  $\pm$   4.568E$-$04\\
  20161031 & $-$3.855E$-$05  $\pm$   7.470E$-$05   &  \phantom{$-$}2.291E$-$04  $\pm$   1.261E$-$04   &  \phantom{$-$}4.941E$-$06  $\pm$   1.759E$-$04   &  \phantom{$-$}8.129E$-$04  $\pm$   4.902E$-$04\\
  20180530 & $-$9.789E$-$05  $\pm$   7.637E$-$05   &  \phantom{$-$}1.044E$-$04  $\pm$   1.278E$-$04   & $-$4.850E$-$04  $\pm$   1.784E$-$04   &  \phantom{$-$}1.443E$-$04  $\pm$   4.982E$-$04\\
\toprule
\end{tabular}
\label{tab:16e}
\end{table*}

\begin{table*}[h]
  \centering
  \caption{Results for the $8\times72$-day sets}
  \scriptsize
\begin{tabular}{lcccc}
\toprule
Start Date &\multicolumn{4}{c}{$\Delta R_{\rm CZ}$/R$_\odot$} \\
of set & $a_2$ & $a_4$ & $a_6$ & $a_8$ \\
\hline
 & \multicolumn{4}{c}{GONG}\\
\hline 
19950718 & $-$2.768E$-$05 $\pm$   1.121E$-$04  &  \phantom{$-$}5.896E$-$07 $\pm$   1.862E$-$04  &  \phantom{$-$}4.227E$-$04 $\pm$   2.477E$-$04  & $-$2.671E$-$04 $\pm$   5.791E$-$04\\
 19960501 & $-$1.740E$-$04 $\pm$   1.078E$-$04  & $-$8.954E$-$04 $\pm$   1.813E$-$04  & $-$9.408E$-$04 $\pm$   2.446E$-$04  & $-$6.814E$-$04 $\pm$   5.705E$-$04\\
 19970213 &  \phantom{$-$}1.695E$-$06 $\pm$   1.036E$-$04  & $-$2.947E$-$04 $\pm$   1.682E$-$04  & $-$6.202E$-$05 $\pm$   2.302E$-$04  & $-$1.127E$-$04 $\pm$   5.302E$-$04\\
 19971128 & $-$2.398E$-$06 $\pm$   1.025E$-$04  &  \phantom{$-$}4.251E$-$04 $\pm$   1.717E$-$04  &  \phantom{$-$}1.401E$-$04 $\pm$   2.354E$-$04  & $-$5.582E$-$04 $\pm$   5.335E$-$04\\
 19980912 & $-$1.036E$-$04 $\pm$   1.042E$-$04  &  \phantom{$-$}5.010E$-$04 $\pm$   1.676E$-$04  & $-$1.571E$-$04 $\pm$   2.347E$-$04  &  \phantom{$-$}2.426E$-$04 $\pm$   5.557E$-$04\\
 19990627 & $-$9.036E$-$05 $\pm$   1.069E$-$04  & $-$7.658E$-$04 $\pm$   1.747E$-$04  & $-$3.574E$-$04 $\pm$   2.406E$-$04  & $-$8.071E$-$04 $\pm$   5.870E$-$04\\
 20000410 & $-$7.886E$-$05 $\pm$   1.030E$-$04  &  \phantom{$-$}3.777E$-$04 $\pm$   1.675E$-$04  & $-$2.984E$-$06 $\pm$   2.337E$-$04  &  \phantom{$-$}2.935E$-$04 $\pm$   5.651E$-$04\\
 20010123 & $-$2.371E$-$04 $\pm$   1.019E$-$04  &  \phantom{$-$}2.336E$-$04 $\pm$   1.668E$-$04  &  \phantom{$-$}2.481E$-$04 $\pm$   2.289E$-$04  &  \phantom{$-$}6.042E$-$04 $\pm$   5.790E$-$04\\
 20011107 &  \phantom{$-$}7.796E$-$05 $\pm$   9.848E$-$05  & $-$3.725E$-$04 $\pm$   1.656E$-$04  & $-$6.139E$-$04 $\pm$   2.302E$-$04  & $-$1.621E$-$05 $\pm$   5.961E$-$04\\
 20020822 &  \phantom{$-$}1.382E$-$04 $\pm$   9.195E$-$05  &  \phantom{$-$}1.558E$-$04 $\pm$   1.596E$-$04  & $-$1.240E$-$04 $\pm$   2.159E$-$04  & $-$9.486E$-$04 $\pm$   5.505E$-$04\\
 20030606 &  \phantom{$-$}4.546E$-$04 $\pm$   9.708E$-$05  & $-$2.729E$-$04 $\pm$   1.606E$-$04  & $-$4.267E$-$05 $\pm$   2.180E$-$04  &  \phantom{$-$}2.217E$-$05 $\pm$   5.547E$-$04\\
 20040320 & $-$5.259E$-$05 $\pm$   9.476E$-$05  & $-$4.382E$-$04 $\pm$   1.549E$-$04  & $-$1.462E$-$03 $\pm$   2.164E$-$04  &  \phantom{$-$}2.752E$-$04 $\pm$   5.265E$-$04\\
 20050102 & $-$3.309E$-$04 $\pm$   9.830E$-$05  & $-$1.548E$-$04 $\pm$   1.611E$-$04  & $-$5.587E$-$04 $\pm$   2.261E$-$04  &  \phantom{$-$}3.355E$-$04 $\pm$   5.552E$-$04\\
 20051017 & $-$2.970E$-$05 $\pm$   1.003E$-$04  &  \phantom{$-$}4.462E$-$04 $\pm$   1.667E$-$04  & $-$7.517E$-$04 $\pm$   2.292E$-$04  &  \phantom{$-$}6.905E$-$04 $\pm$   5.653E$-$04\\
 20060801 & $-$5.539E$-$05 $\pm$   1.009E$-$04  &  \phantom{$-$}4.082E$-$05 $\pm$   1.709E$-$04  & $-$2.453E$-$04 $\pm$   2.309E$-$04  & $-$2.289E$-$04 $\pm$   5.714E$-$04\\
 20070516 & $-$4.547E$-$04 $\pm$   9.964E$-$05  &  \phantom{$-$}2.595E$-$04 $\pm$   1.663E$-$04  & $-$7.486E$-$05 $\pm$   2.259E$-$04  &  \phantom{$-$}1.291E$-$04 $\pm$   5.659E$-$04\\
 20080228 & $-$4.131E$-$04 $\pm$   1.005E$-$04  & $-$8.335E$-$04 $\pm$   1.720E$-$04  &  \phantom{$-$}1.025E$-$03 $\pm$   2.331E$-$04  &  \phantom{$-$}1.958E$-$03 $\pm$   5.736E$-$04\\
 20081212 &  \phantom{$-$}3.793E$-$04 $\pm$   1.002E$-$04  & $-$6.164E$-$04 $\pm$   1.647E$-$04  &  \phantom{$-$}6.337E$-$04 $\pm$   2.261E$-$04  & $-$9.614E$-$04 $\pm$   5.635E$-$04\\
 20090926 &  \phantom{$-$}1.014E$-$04 $\pm$   9.360E$-$05  &  \phantom{$-$}4.171E$-$04 $\pm$   1.573E$-$04  & $-$3.940E$-$04 $\pm$   2.151E$-$04  &  \phantom{$-$}1.164E$-$04 $\pm$   5.285E$-$04\\
 20100711 &  \phantom{$-$}3.666E$-$04 $\pm$   9.888E$-$05  &  \phantom{$-$}1.948E$-$04 $\pm$   1.608E$-$04  &  \phantom{$-$}1.967E$-$05 $\pm$   2.228E$-$04  &  \phantom{$-$}1.424E$-$03 $\pm$   5.518E$-$04\\
 20110425 &  \phantom{$-$}3.511E$-$04 $\pm$   9.564E$-$05  & $-$3.026E$-$04 $\pm$   1.563E$-$04  &  \phantom{$-$}4.268E$-$04 $\pm$   2.187E$-$04  &  \phantom{$-$}4.378E$-$04 $\pm$   5.429E$-$04\\
 20120207 &  \phantom{$-$}2.882E$-$04 $\pm$   9.331E$-$05  &  \phantom{$-$}2.004E$-$04 $\pm$   1.564E$-$04  &  \phantom{$-$}9.971E$-$04 $\pm$   2.156E$-$04  &  \phantom{$-$}8.493E$-$06 $\pm$   5.303E$-$04\\
 20121121 & $-$1.074E$-$04 $\pm$   9.441E$-$05  &  \phantom{$-$}1.453E$-$04 $\pm$   1.572E$-$04  &  \phantom{$-$}2.014E$-$04 $\pm$   2.152E$-$04  &  \phantom{$-$}1.176E$-$04 $\pm$   5.446E$-$04\\
 20130905 & $-$2.840E$-$04 $\pm$   9.456E$-$05  & $-$7.684E$-$04 $\pm$   1.584E$-$04  & $-$3.281E$-$04 $\pm$   2.160E$-$04  & $-$4.680E$-$04 $\pm$   5.513E$-$04\\
 20140620 & $-$1.505E$-$04 $\pm$   9.491E$-$05  & $-$5.927E$-$04 $\pm$   1.566E$-$04  &  \phantom{$-$}3.403E$-$05 $\pm$   2.193E$-$04  &  \phantom{$-$}2.929E$-$04 $\pm$   5.468E$-$04\\
 20150404 & $-$1.182E$-$04 $\pm$   9.221E$-$05  & $-$3.787E$-$04 $\pm$   1.523E$-$04  &  \phantom{$-$}8.066E$-$05 $\pm$   2.127E$-$04  & $-$1.006E$-$03 $\pm$   5.242E$-$04\\
 20160117 &  \phantom{$-$}1.371E$-$04 $\pm$   9.953E$-$05  & $-$2.947E$-$04 $\pm$   1.642E$-$04  & $-$1.361E$-$04 $\pm$   8.816E$-$05  & $-$7.683E$-$04 $\pm$   5.628E$-$04\\
 20161031 & $-$7.077E$-$06 $\pm$   9.976E$-$05  &  \phantom{$-$}2.267E$-$04 $\pm$   1.658E$-$04  &  \phantom{$-$}3.942E$-$04 $\pm$   2.248E$-$04  &  \phantom{$-$}1.029E$-$03 $\pm$   5.638E$-$04\\
 20170815 &  \phantom{$-$}7.155E$-$05 $\pm$   1.019E$-$04  &  \phantom{$-$}6.435E$-$04 $\pm$   1.672E$-$04  &  \phantom{$-$}5.155E$-$04 $\pm$   2.345E$-$04  &  \phantom{$-$}1.014E$-$03 $\pm$   5.626E$-$04\\
 20180530 & $-$1.711E$-$04 $\pm$   1.031E$-$04  &  \phantom{$-$}6.856E$-$04 $\pm$   1.710E$-$04  & $-$2.455E$-$04 $\pm$   2.409E$-$04  &  \phantom{$-$}3.270E$-$06 $\pm$   5.821E$-$04\\
 20190314 & $-$1.832E$-$04 $\pm$   1.026E$-$04  &  \phantom{$-$}5.598E$-$04 $\pm$   1.703E$-$04  & $-$1.281E$-$04 $\pm$   2.407E$-$04  &  \phantom{$-$}2.135E$-$04 $\pm$   5.726E$-$04\\
 20191227 & $-$1.297E$-$04 $\pm$   1.017E$-$04  & $-$2.802E$-$04 $\pm$   1.699E$-$04  & $-$2.404E$-$04 $\pm$   2.363E$-$04  &  \phantom{$-$}8.210E$-$04 $\pm$   5.696E$-$04\\
 20201010 & $-$9.323E$-$05 $\pm$   1.017E$-$04  & $-$5.041E$-$04 $\pm$   1.701E$-$04  &  \phantom{$-$}1.027E$-$04 $\pm$   2.334E$-$04  & $-$7.132E$-$04 $\pm$   5.813E$-$04\\
\hline
 & \multicolumn{4}{c}{MDI}\\
\hline
19960501 & $-$1.679E$-$04 $\pm$   3.243E$-$05  & $-$6.696E$-$04 $\pm$   1.995E$-$04  & $-$1.122E$-$04 $\pm$   2.893E$-$04  &  \phantom{$-$}4.351E$-$04 $\pm$   7.790E$-$04\\
 19970213 & $-$8.866E$-$05 $\pm$   1.185E$-$04  & $-$1.727E$-$04 $\pm$   1.990E$-$04  & $-$6.224E$-$05 $\pm$   2.829E$-$04  & $-$4.827E$-$04 $\pm$   7.841E$-$04\\
 19971128 & $-$2.465E$-$04 $\pm$   1.566E$-$04  &  \phantom{$-$}3.129E$-$04 $\pm$   2.105E$-$04  &  \phantom{$-$}1.275E$-$05 $\pm$   2.979E$-$04  & $-$1.032E$-$03 $\pm$   8.467E$-$04\\
 19980912 & $-$2.922E$-$04 $\pm$   1.198E$-$04  & $-$2.801E$-$04 $\pm$   1.999E$-$04  & $-$8.307E$-$04 $\pm$   2.939E$-$04  & $-$5.026E$-$04 $\pm$   8.066E$-$04\\
 19990627 &  \phantom{$-$}1.646E$-$04 $\pm$   1.166E$-$04  & $-$8.052E$-$04 $\pm$   1.975E$-$04  & $-$8.777E$-$04 $\pm$   2.783E$-$04  & $-$3.687E$-$05 $\pm$   7.673E$-$04\\
 20000410 &  \phantom{$-$}1.817E$-$04 $\pm$   1.176E$-$04  &  \phantom{$-$}4.883E$-$04 $\pm$   1.907E$-$04  &  \phantom{$-$}1.037E$-$04 $\pm$   2.718E$-$04  &  \phantom{$-$}8.150E$-$04 $\pm$   7.486E$-$04\\
 20010123 & $-$1.237E$-$04 $\pm$   1.132E$-$04  &  \phantom{$-$}5.303E$-$04 $\pm$   1.885E$-$04  &  \phantom{$-$}1.721E$-$04 $\pm$   2.701E$-$04  &  \phantom{$-$}9.066E$-$04 $\pm$   7.568E$-$04\\
 20011107 &  \phantom{$-$}1.415E$-$04 $\pm$   1.136E$-$04  & $-$1.612E$-$04 $\pm$   1.935E$-$04  & $-$4.518E$-$04 $\pm$   2.736E$-$04  &  \phantom{$-$}6.942E$-$05 $\pm$   7.623E$-$04\\
 20020822 &  \phantom{$-$}2.985E$-$04 $\pm$   1.139E$-$04  &  \phantom{$-$}1.852E$-$04 $\pm$   1.876E$-$04  & $-$6.602E$-$04 $\pm$   2.712E$-$04  & $-$1.729E$-$03 $\pm$   7.419E$-$04\\
 20030606 &  \phantom{$-$}3.188E$-$04 $\pm$   1.124E$-$04  & $-$3.285E$-$04 $\pm$   1.881E$-$04  & $-$3.241E$-$04 $\pm$   2.704E$-$04  & $-$1.118E$-$03 $\pm$   7.515E$-$04\\
 20040320 &  \phantom{$-$}6.623E$-$05 $\pm$   1.138E$-$04  & $-$3.604E$-$04 $\pm$   1.832E$-$04  & $-$8.670E$-$04 $\pm$   2.714E$-$04  &  \phantom{$-$}1.889E$-$04 $\pm$   7.413E$-$04\\
 20050102 & $-$2.914E$-$04 $\pm$   1.182E$-$04  & $-$1.887E$-$04 $\pm$   1.905E$-$04  & $-$3.699E$-$04 $\pm$   2.837E$-$04  &  \phantom{$-$}5.695E$-$04 $\pm$   7.536E$-$04\\
 20051017 &  \phantom{$-$}8.274E$-$05 $\pm$   1.172E$-$04  &  \phantom{$-$}4.241E$-$04 $\pm$   1.966E$-$04  & $-$7.149E$-$04 $\pm$   2.828E$-$04  &  \phantom{$-$}7.244E$-$04 $\pm$   8.036E$-$04\\
 20060801 & $-$1.746E$-$04 $\pm$   1.259E$-$04  & $-$1.810E$-$04 $\pm$   1.966E$-$04  & $-$5.586E$-$04 $\pm$   2.922E$-$04  &  \phantom{$-$}3.703E$-$05 $\pm$   7.920E$-$04\\
 20070516 & $-$3.081E$-$04 $\pm$   1.235E$-$04  &  \phantom{$-$}4.014E$-$04 $\pm$   2.011E$-$04  & $-$4.883E$-$04 $\pm$   3.002E$-$04  & $-$5.064E$-$04 $\pm$   8.360E$-$04\\
 20080228 & $-$3.176E$-$04 $\pm$   1.261E$-$04  & $-$3.337E$-$04 $\pm$   2.074E$-$04  &  \phantom{$-$}8.664E$-$04 $\pm$   3.022E$-$04  &  \phantom{$-$}6.559E$-$04 $\pm$   8.489E$-$04\\
 20081212 &  \phantom{$-$}3.450E$-$04 $\pm$   1.265E$-$04  & $-$8.905E$-$04 $\pm$   2.091E$-$04  &  \phantom{$-$}3.180E$-$04 $\pm$   3.004E$-$04  & $-$1.078E$-$03 $\pm$   8.400E$-$04\\
\hline
 & \multicolumn{4}{c}{HMI}\\
\hline
  20100711 &  \phantom{$-$}2.960E$-$04 $\pm$   1.005E$-$04  & $-$9.257E$-$05 $\pm$   1.687E$-$04  &  \phantom{$-$}7.059E$-$05 $\pm$   2.342E$-$04  &  \phantom{$-$}6.507E$-$04 $\pm$   6.662E$-$04\\
 20110425 &  \phantom{$-$}3.035E$-$04 $\pm$   9.867E$-$05  & $-$3.076E$-$04 $\pm$   1.622E$-$04  &  \phantom{$-$}6.568E$-$04 $\pm$   2.388E$-$04  &  \phantom{$-$}1.052E$-$03 $\pm$   6.354E$-$04\\
 20120207 &  \phantom{$-$}2.004E$-$04 $\pm$   9.656E$-$05  &  \phantom{$-$}2.827E$-$04 $\pm$   1.613E$-$04  &  \phantom{$-$}1.092E$-$03 $\pm$   2.259E$-$04  &  \phantom{$-$}1.454E$-$04 $\pm$   5.285E$-$05\\
 20121121 &  \phantom{$-$}2.966E$-$05 $\pm$   9.677E$-$05  &  \phantom{$-$}2.004E$-$04 $\pm$   1.602E$-$04  &  \phantom{$-$}4.489E$-$04 $\pm$   2.236E$-$04  & $-$1.435E$-$04 $\pm$   6.175E$-$04\\
 20130905 & $-$1.784E$-$04 $\pm$   9.671E$-$05  & $-$8.528E$-$04 $\pm$   1.614E$-$04  & $-$5.878E$-$04 $\pm$   2.201E$-$04  &  \phantom{$-$}1.844E$-$04 $\pm$   6.199E$-$04\\
 20140620 &  \phantom{$-$}2.003E$-$05 $\pm$   9.364E$-$05  & $-$5.574E$-$04 $\pm$   1.646E$-$04  & $-$8.529E$-$05 $\pm$   2.323E$-$04  &  \phantom{$-$}5.331E$-$04 $\pm$   6.363E$-$04\\
 20150404 & $-$8.478E$-$05 $\pm$   9.577E$-$05  & $-$3.624E$-$04 $\pm$   1.616E$-$04  & $-$1.534E$-$05 $\pm$   2.305E$-$04  & $-$6.690E$-$04 $\pm$   6.253E$-$04\\
 20160117 &  \phantom{$-$}2.328E$-$04 $\pm$   1.001E$-$04  & $-$2.673E$-$04 $\pm$   1.685E$-$04  &  \phantom{$-$}4.211E$-$04 $\pm$   2.341E$-$04  & $-$2.158E$-$04 $\pm$   6.560E$-$04\\
 20161031 &  \phantom{$-$}1.006E$-$04 $\pm$   1.021E$-$04  &  \phantom{$-$}1.180E$-$04 $\pm$   1.744E$-$04  &  \phantom{$-$}2.347E$-$04 $\pm$   2.401E$-$04  &  \phantom{$-$}1.175E$-$03 $\pm$   6.647E$-$04\\
 20170815 &  \phantom{$-$}9.367E$-$05 $\pm$   1.057E$-$04  &  \phantom{$-$}1.250E$-$04 $\pm$   1.735E$-$04  &  \phantom{$-$}3.774E$-$04 $\pm$   2.456E$-$04  &  \phantom{$-$}1.029E$-$03 $\pm$   6.871E$-$04\\
 20180530 & $-$1.398E$-$04 $\pm$   1.079E$-$04  &  \phantom{$-$}5.611E$-$04 $\pm$   1.824E$-$04  & $-$4.173E$-$04 $\pm$   2.494E$-$04  & $-$2.271E$-$04 $\pm$   7.161E$-$04\\
 20190314 & $-$9.100E$-$05 $\pm$   1.120E$-$04  &  \phantom{$-$}4.868E$-$04 $\pm$   1.874E$-$04  & $-$8.267E$-$06 $\pm$   2.726E$-$04  &  \phantom{$-$}8.871E$-$05 $\pm$   7.282E$-$04\\
 20191227 & $-$1.880E$-$04 $\pm$   1.035E$-$04  &  \phantom{$-$}5.511E$-$05 $\pm$   1.711E$-$04  & $-$3.143E$-$04 $\pm$   2.511E$-$04  &  \phantom{$-$}3.282E$-$04 $\pm$   6.994E$-$04\\
 20201010 &  \phantom{$-$}1.053E$-$04 $\pm$   1.014E$-$04  & $-$3.848E$-$04 $\pm$   1.690E$-$04  &  \phantom{$-$}2.648E$-$05 $\pm$   2.373E$-$04  & $-$1.121E$-$03 $\pm$   6.684E$-$04\\
\toprule
\end{tabular}
\label{tab:8e}
\end{table*}


\begin{table*}[h]
  \centering
  \caption{Results for the $5\times72$-day sets}
  \scriptsize
\begin{tabular}{lcccc}
\toprule
Start Date &\multicolumn{4}{c}{$\Delta R_{\rm CZ}$/R$_\odot$} \\
of set & $a_2$ & $a_4$ & $a_6$ & $a_8$ \\
\hline
 & \multicolumn{4}{c}{GONG}\\
\hline 
19960501  & $-$3.588E$-$05            $\pm$   1.319E$-$04    & $-$9.555E$-$04            $\pm$   2.174E$-$04   &  $-$1.704E$-$03              $\pm$   3.081E$-$04    & \phantom{$-$}2.839E$-$04  $\pm$   7.079E$-$04\\
 19970426  & \phantom{$-$}5.235E$-$05  $\pm$   1.254E$-$04    & $-$2.939E$-$04            $\pm$   2.077E$-$04   &  \phantom{$-$}3.811E$-$04  $\pm$   2.894E$-$04    & $-$3.486E$-$04             $\pm$   6.605E$-$04\\
 19980421  & $-$2.454E$-$04            $\pm$   1.224E$-$04    & \phantom{$-$}4.944E$-$04  $\pm$   1.958E$-$04   &  $-$2.373E$-$04              $\pm$   2.715E$-$04    & $-$3.828E$-$04             $\pm$   6.463E$-$04\\
 19990416  & $-$3.982E$-$04            $\pm$   1.248E$-$04    & $-$7.308E$-$05            $\pm$   2.007E$-$04   &  \phantom{$-$}4.238E$-$04  $\pm$   2.924E$-$04    & \phantom{$-$}1.002E$-$03  $\pm$   6.957E$-$04\\
 20000410  & \phantom{$-$}3.672E$-$04  $\pm$   1.254E$-$04    & \phantom{$-$}3.735E$-$04  $\pm$   2.032E$-$04   &  $-$9.070E$-$04              $\pm$   2.842E$-$04    & $-$3.152E$-$04             $\pm$   6.923E$-$04\\
 20010405  & $-$2.928E$-$04            $\pm$   1.199E$-$04    & $-$4.916E$-$05            $\pm$   1.961E$-$04   &  \phantom{$-$}2.728E$-$04  $\pm$   2.752E$-$04    & \phantom{$-$}9.851E$-$04  $\pm$   6.963E$-$04\\
 20020331  & $-$1.104E$-$04            $\pm$   1.232E$-$04    & $-$4.174E$-$04            $\pm$   1.926E$-$04   &  $-$1.592E$-$03              $\pm$   2.699E$-$04    & $-$8.070E$-$04             $\pm$   6.928E$-$04\\
 20030326  & \phantom{$-$}2.632E$-$04  $\pm$   1.083E$-$04    & \phantom{$-$}4.787E$-$04  $\pm$   1.849E$-$04   &  \phantom{$-$}8.017E$-$04  $\pm$   2.524E$-$04    & $-$1.207E$-$03             $\pm$   6.362E$-$04\\
 20040320  & \phantom{$-$}7.737E$-$05  $\pm$   1.132E$-$04    & $-$5.069E$-$04            $\pm$   1.866E$-$04   &  $-$1.369E$-$03              $\pm$   2.646E$-$04    & \phantom{$-$}9.853E$-$04  $\pm$   6.463E$-$04\\
 20050315  & $-$5.549E$-$04            $\pm$   1.141E$-$04    & \phantom{$-$}8.013E$-$04  $\pm$   1.935E$-$04   &  $-$1.379E$-$03              $\pm$   2.636E$-$04    & \phantom{$-$}5.958E$-$04  $\pm$   6.638E$-$04\\
 20060310  & \phantom{$-$}2.420E$-$04  $\pm$   1.196E$-$04    & \phantom{$-$}1.808E$-$04  $\pm$   2.008E$-$04   &  $-$3.567E$-$04              $\pm$   2.713E$-$04    & \phantom{$-$}9.843E$-$04  $\pm$   6.710E$-$04\\
 20070305  & \phantom{$-$}1.442E$-$05  $\pm$   1.206E$-$04    & \phantom{$-$}6.549E$-$04  $\pm$   1.963E$-$04   &  $-$6.800E$-$04              $\pm$   2.784E$-$04    & $-$5.827E$-$04             $\pm$   6.745E$-$04\\
 20080228  & $-$9.281E$-$04            $\pm$   1.246E$-$04    & $-$4.970E$-$04            $\pm$   2.050E$-$04   &  $-$2.762E$-$04              $\pm$   2.874E$-$04    & \phantom{$-$}1.660E$-$03  $\pm$   6.899E$-$04\\
 20090222  & $-$1.614E$-$04            $\pm$   1.193E$-$04    & $-$1.082E$-$03            $\pm$   2.054E$-$04   &  \phantom{$-$}1.009E$-$03  $\pm$   2.777E$-$04    & \phantom{$-$}4.417E$-$04  $\pm$   6.982E$-$04\\
 20100217  & \phantom{$-$}4.107E$-$04  $\pm$   1.185E$-$04    & \phantom{$-$}3.724E$-$04  $\pm$   1.983E$-$04   &  $-$2.199E$-$04              $\pm$   2.767E$-$04    & \phantom{$-$}2.806E$-$04  $\pm$   6.732E$-$04\\
 20110212  & \phantom{$-$}2.066E$-$04  $\pm$   1.129E$-$04    & $-$1.228E$-$04            $\pm$   1.814E$-$04   &  $-$3.439E$-$04              $\pm$   2.537E$-$04    & \phantom{$-$}1.859E$-$03  $\pm$   6.089E$-$04\\
 20120207  & \phantom{$-$}4.621E$-$04  $\pm$   1.126E$-$04    & \phantom{$-$}2.408E$-$05  $\pm$   1.842E$-$04   &  \phantom{$-$}1.372E$-$03  $\pm$   2.652E$-$04    & $-$1.059E$-$04             $\pm$   6.187E$-$05\\
 20130201  & \phantom{$-$}3.435E$-$04  $\pm$   1.105E$-$04    & $-$1.916E$-$04            $\pm$   1.863E$-$04   &  \phantom{$-$}3.637E$-$04  $\pm$   2.588E$-$04    & $-$2.188E$-$04             $\pm$   6.373E$-$04\\
 20140127  & $-$6.314E$-$04            $\pm$   1.101E$-$04    & $-$5.387E$-$04            $\pm$   1.837E$-$04   &  $-$1.264E$-$03              $\pm$   2.585E$-$04    & $-$1.069E$-$03             $\pm$   6.439E$-$04\\
 20150122  & $-$2.463E$-$04            $\pm$   1.116E$-$04    & $-$6.898E$-$04            $\pm$   1.846E$-$04   &  \phantom{$-$}8.648E$-$04  $\pm$   2.553E$-$04    & \phantom{$-$}3.754E$-$04  $\pm$   6.317E$-$04\\
 20160117  & $-$4.519E$-$05            $\pm$   1.129E$-$04    & $-$5.044E$-$05            $\pm$   1.867E$-$04   &  \phantom{$-$}3.817E$-$04  $\pm$   2.592E$-$04    & $-$1.344E$-$03             $\pm$   6.463E$-$04\\
 20170111  & \phantom{$-$}2.313E$-$04  $\pm$   1.156E$-$04    & \phantom{$-$}3.874E$-$04  $\pm$   1.969E$-$04   &  $-$9.711E$-$04              $\pm$   2.686E$-$04    & \phantom{$-$}1.930E$-$03  $\pm$   6.802E$-$04\\
 20180106  & \phantom{$-$}2.746E$-$06  $\pm$   1.232E$-$04    & \phantom{$-$}7.786E$-$04  $\pm$   2.075E$-$04   &  $-$4.994E$-$05              $\pm$   2.776E$-$04    & \phantom{$-$}3.548E$-$04  $\pm$   6.902E$-$04\\
 20190101  & $-$1.264E$-$04            $\pm$   1.227E$-$04    & \phantom{$-$}8.592E$-$04  $\pm$   2.064E$-$04   &  \phantom{$-$}1.348E$-$05  $\pm$   2.897E$-$04    & $-$2.098E$-$03             $\pm$   7.146E$-$04\\
 20191227  & \phantom{$-$}9.851E$-$05  $\pm$   1.184E$-$04    & \phantom{$-$}2.520E$-$04  $\pm$   8.238E$-$05   &  $-$5.347E$-$04              $\pm$   2.786E$-$04    & \phantom{$-$}1.272E$-$03  $\pm$   6.917E$-$04\\
 20201221  & \phantom{$-$}1.857E$-$04  $\pm$   1.194E$-$04    & $-$7.613E$-$04            $\pm$   1.960E$-$04   &  $-$4.214E$-$04              $\pm$   2.781E$-$04    & $-$1.551E$-$04             $\pm$   6.788E$-$04\\
\hline
 & \multicolumn{4}{c}{MDI}\\
\hline
 19960501  & $-$3.249E$-$05            $\pm$   1.496E$-$04    & $-$6.739E$-$04            $\pm$   2.465E$-$04   &  $-$5.510E$-$04              $\pm$   3.652E$-$04    & \phantom{$-$}6.758E$-$04  $\pm$   1.002E$-$03\\
 19970426  & \phantom{$-$}1.408E$-$04  $\pm$   1.413E$-$04    & $-$8.259E$-$05            $\pm$   2.230E$-$04   &  \phantom{$-$}8.475E$-$04  $\pm$   3.428E$-$04    & \phantom{$-$}2.433E$-$04  $\pm$   9.325E$-$04\\
 19980421  & $-$4.957E$-$04            $\pm$   1.669E$-$04    & \phantom{$-$}1.159E$-$03  $\pm$   2.623E$-$04   &  $-$1.588E$-$03              $\pm$   3.901E$-$04    & $-$1.131E$-$03             $\pm$   1.081E$-$03\\
 19990416  & $-$6.330E$-$04            $\pm$   1.397E$-$04    & $-$2.152E$-$04            $\pm$   2.252E$-$04   &  $-$5.035E$-$04              $\pm$   3.297E$-$04    & \phantom{$-$}1.589E$-$03  $\pm$   8.885E$-$04\\
 20000410  & \phantom{$-$}4.040E$-$04  $\pm$   1.355E$-$04    & \phantom{$-$}5.631E$-$05  $\pm$   2.295E$-$04   &  $-$1.165E$-$03              $\pm$   3.199E$-$04    & $-$2.085E$-$04             $\pm$   8.908E$-$04\\
 20010405  & $-$3.047E$-$04            $\pm$   1.399E$-$04    & \phantom{$-$}5.622E$-$04  $\pm$   2.298E$-$04   &  $-$3.100E$-$04              $\pm$   3.347E$-$04    & \phantom{$-$}6.868E$-$04  $\pm$   9.124E$-$04\\
 20020331  & $-$1.789E$-$04            $\pm$   1.372E$-$04    & $-$5.572E$-$04            $\pm$   2.242E$-$04   &  $-$1.441E$-$03              $\pm$   3.222E$-$04    & $-$5.479E$-$04             $\pm$   8.969E$-$04\\
 20030326  & \phantom{$-$}3.194E$-$04  $\pm$   2.223E$-$04    & \phantom{$-$}3.193E$-$04  $\pm$   2.273E$-$04   &  \phantom{$-$}2.665E$-$04  $\pm$   3.130E$-$04    & $-$8.599E$-$04             $\pm$   8.816E$-$04\\
 20040320  & \phantom{$-$}2.402E$-$04  $\pm$   1.355E$-$04    & $-$6.730E$-$04            $\pm$   2.291E$-$04   &  $-$7.335E$-$04              $\pm$   3.338E$-$04    & $-$3.220E$-$04             $\pm$   9.001E$-$04\\
 20050315  & $-$3.596E$-$04            $\pm$   1.345E$-$04    & $-$1.755E$-$04            $\pm$   2.169E$-$04   &  $-$1.810E$-$03              $\pm$   3.311E$-$04    & $-$1.319E$-$04             $\pm$   9.100E$-$04\\
 20060310  & \phantom{$-$}8.495E$-$05  $\pm$   1.465E$-$04    & \phantom{$-$}4.389E$-$05  $\pm$   2.351E$-$04   &  $-$2.662E$-$04              $\pm$   3.444E$-$04    & \phantom{$-$}8.014E$-$04  $\pm$   9.261E$-$04\\
 20070305  & $-$1.004E$-$05            $\pm$   1.484E$-$04    & \phantom{$-$}7.712E$-$04  $\pm$   2.398E$-$04   &  $-$6.664E$-$04              $\pm$   3.555E$-$04    & $-$3.810E$-$04             $\pm$   9.607E$-$04\\
 20080228  & $-$1.021E$-$04            $\pm$   1.549E$-$04    & \phantom{$-$}1.166E$-$04  $\pm$   2.602E$-$04   &  $-$1.027E$-$03              $\pm$   3.755E$-$04    & \phantom{$-$}8.020E$-$04  $\pm$   1.028E$-$03\\
 20090222  & \phantom{$-$}2.454E$-$04  $\pm$   1.513E$-$04    & $-$1.410E$-$04            $\pm$   2.515E$-$04   &  \phantom{$-$}3.212E$-$04  $\pm$   3.520E$-$04    & \phantom{$-$}1.850E$-$04  $\pm$   9.895E$-$04\\
 20100217  & \phantom{$-$}4.135E$-$04  $\pm$   1.466E$-$04    & \phantom{$-$}6.335E$-$04  $\pm$   2.361E$-$04   &  \phantom{$-$}1.217E$-$05  $\pm$   3.362E$-$04    & \phantom{$-$}1.275E$-$04  $\pm$   9.530E$-$04\\
\hline
 & \multicolumn{4}{c}{HMI}\\
\hline
 20110212  & \phantom{$-$}2.221E$-$04  $\pm$   1.158E$-$04    & $-$3.019E$-$04            $\pm$   1.966E$-$04   &  $-$3.283E$-$05              $\pm$   2.749E$-$04    & \phantom{$-$}5.365E$-$04  $\pm$   7.576E$-$04\\
 20120207  & \phantom{$-$}8.597E$-$05  $\pm$   1.199E$-$04    & $-$2.730E$-$04            $\pm$   2.011E$-$04   &  \phantom{$-$}1.516E$-$03  $\pm$   2.815E$-$04    & \phantom{$-$}1.766E$-$04  $\pm$   7.678E$-$04\\
 20130201  & \phantom{$-$}3.320E$-$04  $\pm$   1.133E$-$04    & $-$3.490E$-$04            $\pm$   1.913E$-$04   &  \phantom{$-$}3.782E$-$04  $\pm$   2.728E$-$04    & $-$5.435E$-$04             $\pm$   7.725E$-$04\\
 20140127  & $-$4.768E$-$04            $\pm$   1.144E$-$04    & $-$5.917E$-$04            $\pm$   1.920E$-$04   &  $-$1.209E$-$03              $\pm$   2.717E$-$04    & \phantom{$-$}4.920E$-$04  $\pm$   7.298E$-$04\\
 20150122  & $-$2.282E$-$04            $\pm$   1.143E$-$04    & $-$8.276E$-$04            $\pm$   1.904E$-$04   &  \phantom{$-$}4.310E$-$04  $\pm$   2.738E$-$04    & \phantom{$-$}3.211E$-$04  $\pm$   7.434E$-$04\\
 20160117  & $-$1.630E$-$05            $\pm$   1.192E$-$04    & \phantom{$-$}2.743E$-$05  $\pm$   1.970E$-$04   &  \phantom{$-$}2.686E$-$04  $\pm$   2.685E$-$04    & $-$8.003E$-$04             $\pm$   7.849E$-$04\\
 20170111  & \phantom{$-$}5.495E$-$05  $\pm$   1.165E$-$04    & \phantom{$-$}3.503E$-$04  $\pm$   2.067E$-$04   &  $-$8.565E$-$04              $\pm$   2.853E$-$04    & \phantom{$-$}1.195E$-$03  $\pm$   7.689E$-$04\\
 20180106  & \phantom{$-$}1.701E$-$04  $\pm$   1.305E$-$04    & \phantom{$-$}1.054E$-$04  $\pm$   2.117E$-$04   &  $-$2.592E$-$04              $\pm$   2.929E$-$04    & \phantom{$-$}4.161E$-$04  $\pm$   8.183E$-$04\\
 20190101  & $-$1.138E$-$04            $\pm$   1.274E$-$04    & \phantom{$-$}6.896E$-$04  $\pm$   2.201E$-$04   &  \phantom{$-$}3.924E$-$04  $\pm$   3.210E$-$04    & $-$1.611E$-$03             $\pm$   8.639E$-$04\\
 20191227  & \phantom{$-$}2.661E$-$04  $\pm$   1.332E$-$04    & $-$2.377E$-$04            $\pm$   2.241E$-$04   &  \phantom{$-$}1.229E$-$04  $\pm$   3.164E$-$04    & \phantom{$-$}8.825E$-$04  $\pm$   8.766E$-$04\\
 20201221  & \phantom{$-$}2.669E$-$04  $\pm$   1.245E$-$04    & $-$3.865E$-$04            $\pm$   2.142E$-$04   &  $-$2.735E$-$04              $\pm$   3.101E$-$04    & $-$5.221E$-$04             $\pm$   8.384E$-$04\\
\toprule
\end{tabular}
\label{tab:5e}
\end{table*}

\end{document}